%% file: main.tex
\documentclass[11pt]{article}
\usepackage[margin=1.15in]{geometry}
\usepackage[T1]{fontenc}
\usepackage{lmodern}
\usepackage{amsmath,amssymb,amsthm,mathtools}
\usepackage{booktabs}
\usepackage{graphicx}
\usepackage[normalem]{ulem}
\usepackage[dvipsnames]{xcolor}
\usepackage{tikz}
\usepackage[numbers,sort&compress]{natbib}
\usepackage{microtype}
\usepackage{enumitem}
\usepackage{placeins}
\usepackage{listings}
\lstdefinelanguage{lean4}{
  morekeywords={def,theorem,structure,where,fun,if,then,else,by,namespace,end,open,import,let,have,in},
  sensitive=true, morecomment=[l]{--}, morestring=[b]"}
\newcommand{\leanref}[1]{\texttt{\detokenize{#1}}}
\newcommand{\leantag}[1]{Lean: \leanref{#1}}
\usepackage[font=small,labelfont=bf,labelsep=period,skip=6pt]{caption}
\usepackage[colorlinks=true,linkcolor=MidnightBlue,citecolor=MidnightBlue,urlcolor=MidnightBlue]{hyperref}
\usepackage[capitalise,noabbrev]{cleveref}

\theoremstyle{plain}
\newtheorem{theorem}{Theorem}
\newtheorem{proposition}[theorem]{Proposition}
\newtheorem{lemma}[theorem]{Lemma}

\theoremstyle{definition}
\newtheorem{definition}[theorem]{Definition}
\theoremstyle{remark}
\newtheorem{remark}[theorem]{Remark}
\newtheorem*{question}{Question}

\newcommand{\N}{\mathbb{N}}
\newcommand{\R}{\mathbb{R}}
\newcommand{\Sym}{\mathrm{S}}
\newcommand{\Alt}{\mathrm{A}}
\newcommand{\row}{\mathrm{H}}
\DeclarePairedDelimiter{\abs}{\lvert}{\rvert}
\DeclarePairedDelimiter{\ceil}{\lceil}{\rceil}

\input{numbers.tex}

\title{\bfseries How Unbalanced Must a Latin Rectangle Be?}
\author{Zhengyu Li\thanks{Georgia Institute of Technology. \texttt{brian.li@gatech.edu}}
  \and Vijay Ganesh\thanks{Georgia Institute of Technology. \texttt{vganesh45@gatech.edu}}}
\date{}

\begin{document}
\maketitle

\begin{abstract}
\noindent
We view a $k\times n$ Latin rectangle as $n$ points in the grid $[n]^k$. Each symbol gives one
point, and its coordinates are the columns that the symbol occupies. The rectangle is
\emph{balanced} when these points are pairwise equidistant in the $\ell_1$ metric. Its
\emph{imbalance} measures how far the rectangle is from this ideal. We study the minimum imbalance from three sides.
\emph{Constructively}, we use a simulated annealing that moves symbols within rows and only
penalises column clashes. Against the reference table of Voboril et al.\ (2025), it improves
\NumImproved{} of the \NumCells{} non-square values for $8\le n\le 13$, by up to \MaxGain\%, and
matches the rest. A cheap invariant, the \emph{row group}, certifies that none of the 33 new
rectangles is isotopic to rows of a group table. Hence the earlier column-swapping local search
cannot reach any of them.
\emph{Exactly}, a computer-assisted branch and bound proves that the minimum imbalance of a
$3\times 9$ rectangle is $56$. Díaz et al.\ found this value, but its optimality was open. A sharper
pruning bound, derived from the identity $I=2\sum(d-t)^+$, reduces the search tree by a factor of
$2.4$ to $3.3$ on the instances with $k\ge3$ that we solve.
\emph{Asymptotically}, the minimum imbalance is $\Theta(n^3)$ for every fixed $k$. A pigeonhole
argument gives a lower constant of order $k\,4^{-k}$, and random rectangles give an upper constant
of order $\sqrt k$. For $k=2$, the limit of $\min I/n^3$ exists, a permuton attains it, and it lies
between $4/375$ and $0.1196$. We compute the upper value exactly from one explicit permuton. Perfect balance is an $\ell_1$-equilateral problem, so
it forces $k=\Omega(n/\log n)$ rows.
For $2\times n$ rectangles, single two-minute annealing runs improve five of the seven six-hour MIP
upper bounds for $14\le n\le 20$.
Finally, the kernel of the Lean~4 proof assistant checks every rectangle reported in this paper
against a 74-line specification that imports only Lean's core library. We also prove the pigeonhole
lower bound in Lean with Mathlib.
\end{abstract}

\section{Introduction}

In an agricultural field trial, the treatments occupy a grid of plots. When some pairs of
treatments are systematically planted closer together than others, spatial correlation in the
soil can bias their comparison~\citep{vanEs1993}. Latin designs remove the row and column
effects, but not this one. \emph{Spatially balanced} Latin squares ask, in addition, that every
pair of treatments be equally far apart on average~\citep{GomesSellmann2004,LeBrasGomesSelman2012}.
\citet{DiazLeBrasGomes2017} posed the same question for $k\times n$ rectangles, observed that
perfect balance is then usually impossible, and asked instead for rectangles of minimum
\emph{imbalance}. Later work on the problem uses SAT encodings that prove optima with checkable proof
certificates~\citep{RamaswamySzeider2023} and mixed-integer programming~\citep{MirkaEtAl2023}. It also
uses streamliners written by large language models~\citep{VoborilRamaswamySzeider2025} and LLM agents
that reformulate models~\citep{VoborilSzeider2026}. Closest to our work, \citet{XiaEtAl2026} use
simulated annealing for spatially balanced Latin \emph{squares} and reach the optimum for $n=10$ and
$n=13$ in under a second. They also prove an $\Omega(n^2)$ lower bound on the imbalance of squares
with $n\equiv1\pmod 3$, by a parity argument. The starting point of this paper is the reference table
of~\citet{VoborilRamaswamySzeider2025}, which collects the best values that the methods above obtained
for rectangles. To our knowledge, this table is still current. The later
work~\citep{VoborilSzeider2026} reports aggregate model comparisons but no per-instance values, and
\citet{XiaEtAl2026} treat only squares.

Here is the point of view of this paper. Record, for each symbol $u$, the columns
$c_1(u),\dots,c_k(u)$ that it occupies in the $k$ rows. This gives a point $x_u\in[n]^k$. The distance between two symbols that the literature uses is
exactly the $\ell_1$ distance $\lVert x_u-x_v\rVert_1$. A balanced rectangle is therefore a set of $n$
points in a $k$-dimensional grid that are pairwise \emph{equidistant} in the taxicab metric.
When $k$ is small compared with $n$, the grid is too crowded for equidistance, and
\cref{sec:cubic} makes this quantitative. When $k=n$ there is room, consistent with the existence
of balanced squares for infinitely many $n$. The point picture also suggests a search that moves
individual points, rather than one that rearranges a rigid algebraic pattern.

\paragraph{Contributions.}
\begin{enumerate}[leftmargin=*,itemsep=2pt]
\item \textbf{New upper bounds} (\cref{sec:search,sec:results}). The search keeps each row a permutation and turns the column condition into a penalty with an
adaptive weight. A move swaps two symbols within a row, and the search evaluates its effect exactly in
$O(n)$ time. Against the reference table of
\citet{VoborilRamaswamySzeider2025}, the search improves \NumImproved{} of \NumCells{}
instances and matches the other \NumMatched{}.

\item \textbf{A new optimum} (\cref{sec:exact}). A branch and bound that places one symbol at a time proves that the known value $56$ is optimal for
$3\times 9$, which was open. It also independently reproduces the known optima for $2\times n$ with
$n\le 11$ and for $3\times 8$ and $4\times 8$.
A second, independent implementation with a sharper bound confirms the $3\times9$, $3\times8$
and $4\times8$ results and the $2\times10$ and $2\times11$ optima.

\item \textbf{A certificate of non-group structure} (\cref{sec:orbit}). The permutation group generated by the ratios of the rows does not change under column operations.
For rectangles taken from group tables, its order divides $n$. For 32 of our 33 improved rectangles it is the full
alternating or symmetric group, and for the last one it has order $46080$. So local search that only permutes the columns of a cyclic rectangle~\citep{DiazLeBrasGomes2017}
cannot reach any of them. The invariant does not separate our rectangles from the
published SAT and streamliner rectangles, most of which are also far from group tables.

\item \textbf{Cubic growth} (\cref{sec:cubic}). For every fixed $k$, the minimum imbalance is
$\Theta(n^3)$. The lower constant $c_k=\max_{m\ge3} k\,m^{-k}(\frac13-\frac1m)$ is of order $k\,4^{-k}$,
while random rectangles give an upper constant $D_k\le\sqrt k/(6\sqrt2)$. The right order in $k$
is open. Perfect balance also requires $n\le e(\ell_1^k)=O(k\log k)$, by a classical bound on
equilateral sets.

\item \textbf{The case $k=2$} (\cref{sec:two}). On the open instances $14\le n\le 20$, we compare a single-seed two-minute annealing run with the
six-hour upper bounds of \citet{MirkaEtAl2023}. The run improves the bound for $15\le n\le19$, ties at
$n=14$ and is worse at $n=20$. The limit of $\min I/n^3$ exists and is the minimum of a continuous
functional on permutons. We bracket it between $4/375$ and $0.1196$, and we compute the upper value
exactly for the permuton of one $2\times64$ rectangle.

\item \textbf{A Lean formalisation} (\cref{sec:lean}). Lean's kernel certifies all 62 rectangles behind our upper bounds by evaluation. The check uses no \texttt{native\_decide} and no
Mathlib, only Lean's core library, so it trusts only the kernel and a short specification. We prove the identity~\eqref{eq:half} and \cref{thm:cubic} in Lean with Mathlib. Each formalised statement
names the Lean theorem that proves it.
\end{enumerate}

None of the algorithmic ingredients is new by itself. Simulated annealing~\citep{KirkpatrickGelattVecchi1983} is established, including for the square
version of this problem~\citep{XiaEtAl2026}. So are row-preserving swaps with temporarily violated
column constraints~\citep{ZhangMa2008,XieEtAl2024} and adaptive penalties, also known as strategic
oscillation~\citep{GloverHao2011}. What we add is their
combination for rectangles, the resulting records, and the exact and asymptotic results.

\section{Preliminaries}\label{sec:prelim}

A $k\times n$ \emph{Latin rectangle} ($k\le n$) is an array with entries in
$[n]=\{1,\dots,n\}$ in which no symbol repeats in a row or in a column. Let $c_i(u)$ denote the
column of symbol $u$ in row $i$ and $x_u=(c_1(u),\dots,c_k(u))\in[n]^k$. Following
\citet{DiazLeBrasGomes2017} and \citet{RamaswamySzeider2023},
\[
  d(u,v)=\sum_{i=1}^k\abs{c_i(u)-c_i(v)}=\lVert x_u-x_v\rVert_1,
  \qquad
  I(L)=\sum_{1\le u<v\le n}\abs[\big]{d(u,v)-t},
  \qquad t=\frac{k(n+1)}{3}.
\]
Each row contributes $\sum_{u<v}\abs{c_i(u)-c_i(v)}=\binom{n+1}{3}$ to the total distance, so
$t$ is the \emph{average} distance over all $\binom n2$ pairs. In particular the deviations
$d(u,v)-t$ sum to zero, and
\begin{equation}\label{eq:half}
  I(L)=2\sum_{u<v}\bigl(t-d(u,v)\bigr)^+=2\sum_{u<v}\bigl(d(u,v)-t\bigr)^+ .
\end{equation}
The pairs that are too close and the pairs that are too far therefore contribute equally to the
imbalance. This identity drives both the pruning bound of \cref{sec:exact} and the lower bound of
\cref{sec:cubic}. (In Lean, the average is \leanref{LatinRect.sum_dist} and the identity is
\leanref{LatinRect.imbalance_eq_two_mul_sum_pos}.) The quantity $3I$ is an integer. As in the literature, we write imbalances as $x.0$, $x.3$ and
$x.6$ for $x$, $x+\frac13$ and $x+\frac23$.

\paragraph{Two dictionaries.} We can read a rectangle through its rows or through its points.
Through its rows it is a $k$-tuple $(r_1,\dots,r_k)$ of bijections $r_i\colon[n]\to[n]$ from
columns to symbols, with $r_i(j)\ne r_{i'}(j)$ for $i\ne i'$. Through its points it is a set $\{x_u\}$ in $[n]^k$ whose coordinate projections are bijections.
The column condition says exactly that every point has distinct coordinates.
The imbalance is invariant under relabelling symbols, permuting rows and reversing the order of
the columns, $j\mapsto n+1-j$, which preserves every distance. It is not invariant under general
column permutations. After relabelling the symbols, we may assume $r_1=\mathrm{id}$.

\begin{figure}[t]
\centering
\includegraphics[width=\linewidth]{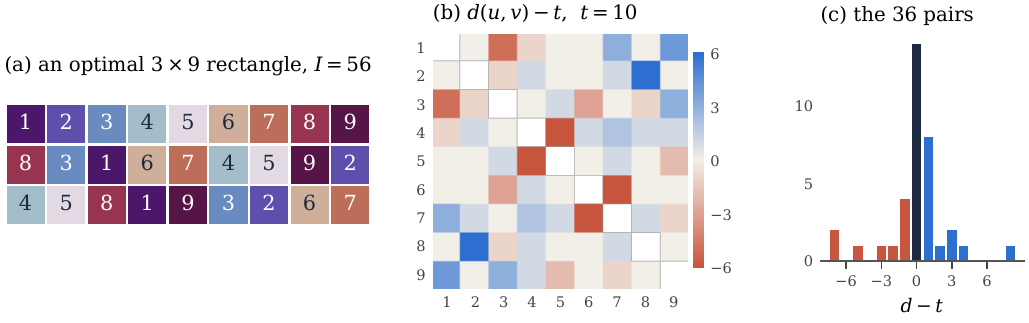}
\caption{A $3\times 9$ rectangle of imbalance $56$, which \cref{thm:3x9} proves optimal.
(a)~The rectangle, one colour per symbol. (b)~The deviation $d(u,v)-t$ of every pair from the
average distance $t=10$. Blue pairs are too far apart and red pairs too close. (c)~Histogram of
the $36$ deviations. They sum to zero, and $I=\sum\abs{d-t}=56$.}
\label{fig:optimum}
\end{figure}

\section{Search: rows as permutations, columns as a penalty}\label{sec:search}

The state of the search is a $k$-tuple of permutations with $r_1=\mathrm{id}$. It need not be
Latin. A \emph{move} picks a row $i\ge2$ and two columns $j_1\ne j_2$, and swaps the symbols
$a=r_i(j_1)$ and $b=r_i(j_2)$. In the point picture, the move exchanges the $i$-th coordinates of
the two points $x_a$ and $x_b$, and nothing else moves.

\begin{lemma}[Incremental evaluation]\label{lem:delta}
After the move, only the $2(n-2)$ distances $d(a,w)$ and $d(b,w)$ with $w\notin\{a,b\}$ change,
and with $p=c_i(w)$,
\[
  d'(a,w)=d(a,w)-\abs{j_1-p}+\abs{j_2-p},\qquad
  d'(b,w)=d(b,w)-\abs{j_2-p}+\abs{j_1-p}.
\]
Hence the exact change of $3I$ is computable in $O(n)$ integer operations.
\end{lemma}

\begin{proof}
The move only exchanges the $i$-th coordinates of $x_a$ and $x_b$. So $d(a,b)$ does not change,
and no pair without $a$ or $b$ is affected.
\end{proof}

\paragraph{Penalty.} Let $V$ be the number of excess occurrences of symbols in columns. Per-column counters maintain $V$ in $O(1)$ time per move. The search minimises $3I+W\cdot V$ and
records only states with $V=0$. The weight starts at $W=k$ and stays in $[1,4n^2]$. Every $2^{16}$ proposals, the search multiplies
$W$ by $1.02$ if the current state has conflicts and by $0.99$ otherwise. The search can therefore
leave the Latin rectangles, pass through non-Latin states and return. By contrast, moves that permute whole columns never leave the Latin
rectangles, and as \cref{sec:orbit} shows, they never change the row group.

\paragraph{Schedule.} The search starts from the cyclic rectangle $r_i(j)=i+j \bmod n$. It accepts
a move of change $\Delta$ with probability $\min(1,e^{-\Delta/T})$. The temperature starts at $T_0=2k$ (in units of $3I$) and is multiplied by $0.9999995$ per
proposal. It never drops below $0.05$, which it reaches after about $10^7$ proposals. Most of a run is therefore a
low-temperature walk on plateaus. If the best Latin rectangle found so far does not improve for $4\cdot10^7$ proposals, the search
resets the temperature to $T_0$. The search discards a proposal with $j_1=j_2$ but still counts it,
and does not cool for it. Runs are single-threaded, and the search re-evaluates the reported
rectangle from scratch.

An independent checker, written from the definition with exact rational arithmetic, verifies every
reported rectangle. It also reproduces the imbalances reported for all fourteen rectangles printed
in~\citep{RamaswamySzeider2023,VoborilRamaswamySzeider2025}. Lean's kernel also certifies every
reported rectangle (\cref{sec:lean}).

\section{Improved upper bounds}\label{sec:results}

We take the \NumCells{} non-square instances $8\le n\le13$, $2\le k\le n-1$ of Table~1
of~\citep{VoborilRamaswamySzeider2025}, which also collects the earlier records. All runs used an
Apple M3 laptop, one core per run. Each instance received independent runs of 1 and 5 CPU minutes. The larger instances ($k\ge4$,
$n\ge9$) also received a 10-minute run, so each instance used at most 16 CPU minutes. \Cref{tab:results} reports the best of these runs. Of the
\NumImproved{} new values, 10 came first from the 1-minute run, 8 from the 5-minute run and 15
from the 10-minute run.
The search improves \NumImproved{} values, matches \NumMatched{} and is worse on none. The mean
improvement is \MeanGain\% and the largest is \MaxGain\% (\TopCell: \TopOld{}$\,\to\,$\TopNew). The
search matches every record with $n=8$ or $k=2$, every record with $k=3$ except $3\times12$, and
the records for $4\times10$, $5\times10$ and $6\times12$. Several of these are proven
optimal~\citep{RamaswamySzeider2023,MirkaEtAl2023}, so matching is the best possible there.

\begin{table}[t]
\centering
\small
\input{table.tex}
\caption{Minimum known imbalance of $k\times n$ Latin rectangles. Bold: new value from this
work (best of the 1-, 5- and 10-minute runs), with the previous best value struck through below it.
Previous values are those collected in Table~1 of~\citet{VoborilRamaswamySzeider2025}, which includes results of
\citep{DiazLeBrasGomes2017,RamaswamySzeider2023,MirkaEtAl2023}, of 4-hour Gurobi runs and of
LLM-generated streamliners. Plain: a record that our search matches. Every bold or plain value is attained by a
rectangle certified in Lean (\leanref{Certificates.exists_k*_n*}, \cref{sec:lean}). The
struck-through values are taken from the literature and are not certified.}
\label{tab:results}
\end{table}

\paragraph{Variability.} Single runs are noisy. In 9 of the \NumImproved{} improved cells, a longer run ended with a worse value than a shorter
one. Across the three budgets, the final values of a cell differ by a median of 5\% and at most
32.5\%. The values in \cref{tab:results} are
therefore upper bounds from a few runs, not estimates of typical performance.

\paragraph{Prior effort.} Most previous values come from \citet{VoborilRamaswamySzeider2025}. They
add LLM-generated streamlining constraints to a MiniZinc model and solve it with Gurobi. They run the
streamliners either incrementally, warm-starting each round from the best solution so far, or with an
evolutionary search. Their runs take 4 hours or more on dual 10-core Xeon nodes. The other previous
values come from 4-hour Gurobi runs and earlier
literature~\citep{DiazLeBrasGomes2017,RamaswamySzeider2023,MirkaEtAl2023}.

These runs used different hardware, different time limits and a general-purpose solver, so we
compare values, not methods. Several previous values are attested by printed rectangles. For
instance, the value $274.0$ for $12\times13$ is the rectangle in Fig.~6h
of~\citep{VoborilRamaswamySzeider2025}, and our checker and Lean reproduce it. The amount of search
behind each previous value also varies between cells. So the size of an improvement need not reflect
the difficulty of an instance.

\begin{figure}[t]
\centering
\includegraphics[width=\linewidth]{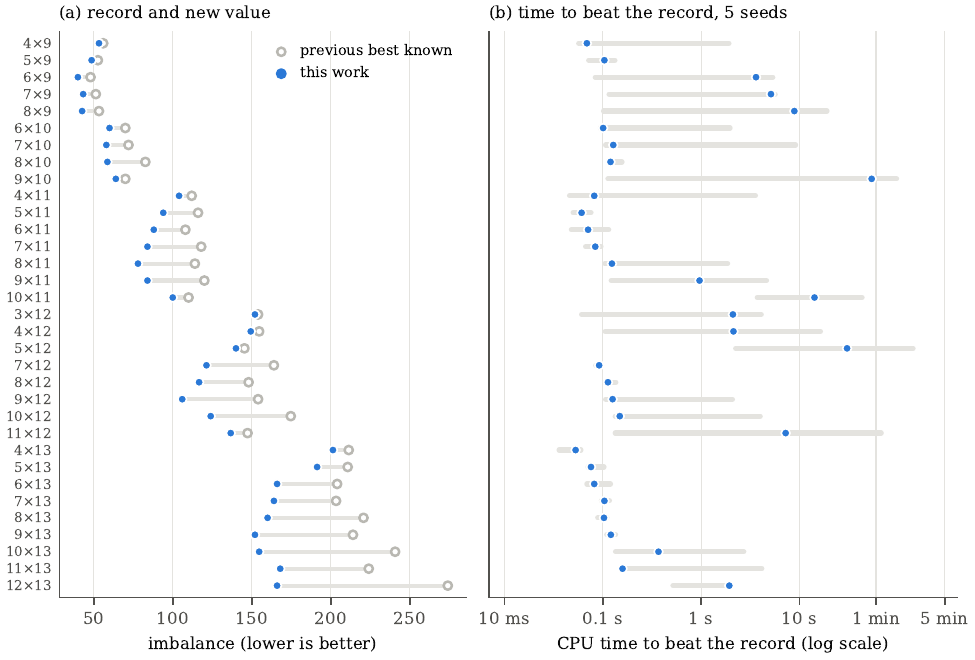}
\caption{The \NumImproved{} improved instances. (a)~Previous record (open circle) and new value
(filled). (b)~CPU time until the search first beats the record: median (dot) and range (bar)
over \NumSeeds{} seeds, log scale.}
\label{fig:records}
\end{figure}

\paragraph{Speed.} We reran the search with \NumSeeds{} fresh seeds on every improved instance,
stopping when it beat the record (\cref{fig:records}b). All \NumTimingRuns{} runs succeeded. The
median time to beat the previous record (not to reach our final value) was \MedianBeat{} CPU
seconds and the maximum \MaxBeat{} seconds. For the reasons above we make no runtime comparison
with earlier work. We measured these times on the Apple M3 laptop and did not archive their logs. An independent
rerun of the same \NumTimingRuns{} runs used the original source on Linux x86-64 with GCC. It beat
every record (pooled median $0.21$ CPU seconds, maximum $280$) and ended at the same values as the
original runs. Its logs are in the \texttt{validation/} directory of the repository.

\section{Exact search and a new optimum}\label{sec:exact}

To complement the upper bounds, we wrote a branch and bound that decides whether a rectangle
with $3I<\beta$ exists for a given bound $\beta$.

\paragraph{Branching.} With $r_1=\mathrm{id}$ fixed, the search places the symbols in the order $1,2,\dots,n$.
Placing a symbol means choosing its column in each row $2,\dots,k$, subject to the Latin
conditions. The search places symbol $1$ in increasing columns of rows $2,\dots,k$. This loses nothing.
Relabelling the symbols makes $r_1=\mathrm{id}$, and permuting rows $2,\dots,k$ keeps
$r_1=\mathrm{id}$. Symbol $1$ occupies distinct columns in those rows, so exactly one such
permutation makes its columns increase with the row index. Both operations
preserve $I$. A \emph{node} is one choice of a column for a symbol in one row that passes the Latin
check. When symbol $u$ is complete, the deviations $\abs{3d(u,v)-3t}$ for all $v<u$ are final, and the
search adds them to the cost. We do not yet exploit the choice of the row that is fixed to the
identity, or the column reversal of \cref{sec:prelim}. Together they offer up to another factor $2k$
of symmetry.

\paragraph{Bounds.} Every unplaced row adds at least $1$ to any distance, because the two symbols
occupy different columns, and at most $n-1$. So, while the search places a symbol, each pair
$(u,v)$ with $v<u$ has a known \emph{overshoot} $o_{uv}=\bigl(3\,d_{\mathrm{part}}+3\rho-3t\bigr)^+$
and a known \emph{shortfall} $s_{uv}=\bigl(3t-3\,d_{\mathrm{part}}-3\rho(n-1)\bigr)^+$. Here $d_{\mathrm{part}}$ is the distance over the
rows placed so far, and $\rho$ is the number of rows still to place. Let $P$ and $N$ be the sums of the positive and negative parts of the final deviations
$3d-3t$, so that the cost so far is $P+N$. By~\eqref{eq:half}, $3I=2\sum(3d-3t)^+=2\sum(3d-3t)^-$,
so each of
\[
  P+N+\sum o_{uv}+\sum s_{uv},\qquad 2\Bigl(P+\sum o_{uv}\Bigr),\qquad 2\Bigl(N+\sum s_{uv}\Bigr)
\]
is a lower bound on $3I$. The search prunes the node when their maximum reaches $\beta$. The first
version of our program used only $P+N+\sum o_{uv}$. The other two terms cost a few additions per
node. When a symbol is complete, the first program also adds
$m_0=\min(3t\bmod 3,\,3-3t\bmod 3)$ for every pair that is not yet final. This is valid only for the
first of the three bounds, because a pair's deviation can lie on either side of $t$. For
$3\times n$ and $4\times 8$ we have $m_0=0$.
A run with $\beta$ equal to three times the value of our rectangle either finds a better rectangle
or proves that none exists.

\begin{theorem}[computer-assisted]\label{thm:3x9}
The minimum imbalance of a $3\times 9$ Latin rectangle is $56$.
\end{theorem}

\noindent The upper bound is the rectangle of \cref{fig:optimum}, certified in Lean as \leanref{Paper.three_by_nine}. The lower bound is not formalised
(\cref{sec:lean}). With the first bound, the branch
and bound explores $5.5\cdot10^9$ nodes in $140$ CPU seconds and finds no rectangle with $I<56$.
As a sanity check, a run with $\beta=169$ finds a rectangle of value $56$. \citet{DiazLeBrasGomes2017}
had already found the value $56$ itself. To our knowledge, its optimality was not known. That paper
does not prove it, and Table~1 of~\citep{VoborilRamaswamySzeider2025} lists it as unproven.

\paragraph{A second implementation.} To reduce the dependence on a single program, we re-derived the result with a second
implementation. It was written independently from the description above and uses the sharper bound. Restricted to the first bound, it explores the same number of nodes as the first program to
the precision we recorded ($7.77\cdot10^7$ against $7.8\cdot10^7$ for $3\times 8$, $5.48\cdot10^9$
against $5.5\cdot10^9$ for $3\times9$). With the sharper bound, it confirms the three results with $k\ge3$ using $2.4$ to $3.3$ times fewer
nodes. It confirms the $2\times10$ and $2\times11$ optima with about $1.5$ times fewer nodes
(\cref{tab:exact}). Both programs share the branching scheme, so this guards against
coding errors but not against an error in the scheme itself.

\begin{table}[t]
\centering
\footnotesize
\setlength{\tabcolsep}{4pt}
\begin{tabular}{@{}lrrrrl@{}}
\toprule
instance & optimum $I$ & nodes, first bound & CPU time & nodes, sharper & before this work \\
\midrule
$2\times n$, $n\le 11$ & as known & $\le 3.8\cdot10^7$ & $\le 1.1$\,s & $\le 2.6\cdot10^7$ ($n\ge10$) & proven \citep{DiazLeBrasGomes2017,MirkaEtAl2023} \\
$3\times 8$ & 36.0 & $7.8\cdot10^7$ & 1.9\,s & $3.2\cdot10^7$ & proven \citep{DiazLeBrasGomes2017,RamaswamySzeider2023} \\
$4\times 8$ & 32.0 & $1.9\cdot10^{10}$ & 477\,s & $5.7\cdot10^9$ & proven \citep{VoborilRamaswamySzeider2025} \\
$3\times 9$ & \textbf{56.0} & $5.5\cdot10^9$ & 140\,s & $2.1\cdot10^9$ & \textbf{open} \\
\midrule
$5,6,7\times 8$;\ $4$--$8\times 9$ & --- & --- & $>1800$\,s & --- & open (undecided) \\
\bottomrule
\end{tabular}
\caption{Exact branch and bound, single core. Each run proves that no rectangle beats our upper
bound. The sharper-bound node counts are from the second implementation. We omit its CPU times because
they were measured on a shared machine. On all small instances with $n\le7$ and $k\le4$ the first
program also reproduces the values reported in~\citep{DiazLeBrasGomes2017}.}
\label{tab:exact}
\end{table}

\Cref{tab:exact} summarises the runs. The method is deliberately plain: no symmetry beyond row
order and symbol relabelling, and no learned bounds. With a 30-minute cap per instance and the first
bound, the remaining open instances with $n\in\{8,9\}$ stay undecided. Unlike
the DRAT proofs of~\citep{RamaswamySzeider2023}, our search does not emit a checkable
certificate. \Cref{thm:3x9} therefore rests on the correctness of two short programs. The
first, which we publish, reproduces all the known optima listed above. The second, which is not yet in the repository, reproduces those for $2\times10$, $2\times11$, $3\times8$ and $4\times8$. A certificate in the style
of~\citep{RamaswamySzeider2023} is the natural next step, and the overshoot and shortfall bounds
may be worth encoding there as redundant constraints.

\section{A certificate that the new rectangles are not group-based}\label{sec:orbit}

\citet{DiazLeBrasGomes2017} start their local search from $k$ random rows of the cyclic Latin square.
They repeatedly swap two random columns, keep a swap only when it lowers the imbalance, and restart
with fresh rows. They note that the method \textquotedblleft never leaves its isotopic class\textquotedblright. Two rectangles are \emph{isotopic} if permuting rows, columns and symbols turns one into the other.
So permuting columns preserves the isotopy class by definition. The question is how to see that a given rectangle lies
outside every such class. We give an invariant that is cheap to compute and detects whether a
rectangle can come from a group table. It is not a complete isotopy invariant, and it does not by
itself explain why one search outperforms another.

\begin{definition}
The \emph{row group} of a Latin rectangle with rows $r_1,\dots,r_k$ is the permutation group on
symbols
\[
  \row(L)=\bigl\langle\, r_i\circ r_j^{-1} : 1\le i,j\le k \,\bigr\rangle\ \le\ \Sym_n .
\]
\end{definition}

The element $r_i\circ r_j^{-1}$ sends the symbol in column $c$ of row $j$ to the symbol in column
$c$ of row $i$. Since $r_i r_j^{-1}=(r_ir_1^{-1})(r_jr_1^{-1})^{-1}$, the group is already
generated by the $k-1$ ratios $r_ir_1^{-1}$. In particular, for $k=2$ the row group is the cyclic
group generated by $r_2r_1^{-1}$, so it carries no information beyond the cycle type of one
permutation.

\begin{proposition}\label{prop:rowgroup}
\leavevmode
\begin{enumerate}[label=(\roman*),itemsep=1pt]
\item Permuting columns does not change $\row(L)$. Permuting rows permutes its generators.
Relabelling symbols by $\sigma$ replaces it by $\sigma\row(L)\sigma^{-1}$. So the conjugacy
class of $\row(L)$ is an isotopy invariant.
\item If $L$ is isotopic to $k$ rows of the Cayley table of a group $G$ of order $n$, then
$\row(L)$ is conjugate to a subgroup of the left-regular representation of $G$. In particular
it acts semiregularly (only the identity fixes a symbol) and $\abs{\row(L)}$ divides $n$.
\item For $i\ne1$, a swap in row $i$ left-multiplies $r_ir_1^{-1}$ by a transposition and fixes
the other generators. A single swap can therefore enlarge the row group.
\end{enumerate}
\end{proposition}

\begin{proof}
(i)~Replacing every row by $r_i\circ\pi$ gives $(r_i\pi)(r_j\pi)^{-1}=r_ir_j^{-1}$. The other
two statements are immediate. (ii)~Up to isotopy, $r_i(c)=\sigma(a_i b_c)$ with $a_i\in G$ and
$c\mapsto b_c$ a bijection onto $G$. Then $r_ir_j^{-1}$ maps $\sigma(a_jb)$ to $\sigma(a_ib)$,
so it is $\sigma\lambda_{a_ia_j^{-1}}\sigma^{-1}$, where $\lambda_g$ is left multiplication by
$g$. (iii)~Swapping two symbols in row $i$ replaces $r_i$ by $\tau r_i$ for a transposition
$\tau$.
\end{proof}

\begin{figure}[t]
\centering
\includegraphics[width=0.92\linewidth]{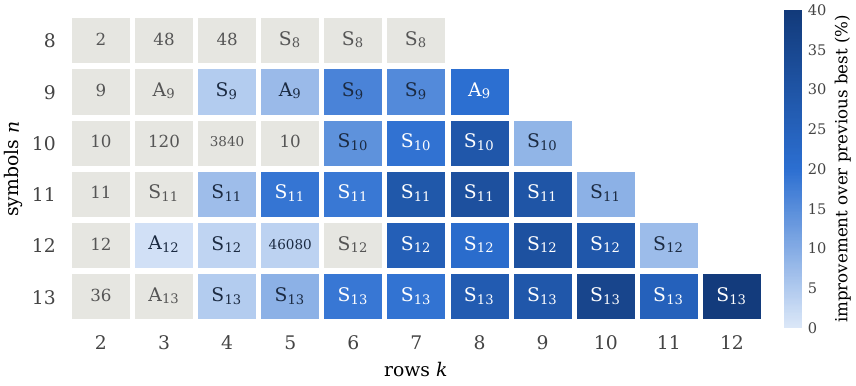}
\caption{Row groups of our best rectangles in the benchmark range. Shading shows the improvement
over the previous best value, and grey cells are matched records. Each label gives the row group: $\Sym_n$ or
$\Alt_n$, otherwise its order. For $k=2$ the row group is cyclic, generated by one permutation. It still certifies
non-group structure when its order does not divide $n$.}
\label{fig:rowgroups}
\end{figure}

So any search whose moves permute columns, starting from rows of a cyclic Latin square, stays
among rectangles whose row group is cyclic of order dividing $n$. \Cref{fig:rowgroups} shows the row groups of the best rectangles we
found. Of the \NumImproved{} improved rectangles, 29 have row group $\Sym_n$ and 3 have $\Alt_n$
($5\times9$, $8\times9$ and $3\times12$). The remaining one, $5\times 12$, has a row group of order
$46080=2^6\cdot 6!$, which does not divide $12$. By \cref{prop:rowgroup}(ii), none of the 33 rectangles is isotopic to rows of a group table. In particular, the column-swapping
search of~\citep{DiazLeBrasGomes2017} cannot reach any of them. The row group of the $5\times12$
rectangle is the full centraliser $\Sym_2\wr\Sym_6$ of the pairing $u\leftrightarrow 13-u$ of symbols.
This mirror symmetry also appears in the published $4\times12$ and $5\times12$
records.

\paragraph{What this does and does not show.} The row group certifies that column-based local search cannot reach our rectangles. It does not
explain why annealing finds better upper bounds than those reported by the exact solvers and
streamliners, which supply most of the records we improve. Ten of the fourteen published SAT and
streamliner rectangles also have row group $\Sym_n$. The SAT encodings aim at optimality proofs,
while the streamliner approach searches a general-purpose model with a MIP solver under LLM-proposed
restrictions. We have not isolated which ingredient of our search accounts for the better upper bounds.

\FloatBarrier
\section{Cubic growth for a fixed number of rows}\label{sec:cubic}

Why does imbalance grow so quickly with $n$ (\cref{tab:results}), and why can squares be perfectly
balanced? The point picture gives a short answer. For fixed $k$, there is not enough room in
$[n]^k$ for $n$ points that are pairwise far apart.

\begin{theorem}[\leantag{cubic_lower_bound_ceil}]\label{thm:cubic}
Let $L$ be a $k\times n$ Latin rectangle and $s\ge 1$ an integer, and put
$M=\ceil{n/s}$. Then
\[
  I(L)\ \ge\ n\Bigl(\frac n{M^k}-1\Bigr)\Bigl(\frac{k(n+1)}3-k(s-1)\Bigr)
\]
whenever $k(s-1)\le t$, that is, $3(s-1)\le n+1$. Consequently, for every fixed $k$,
\[
  c_k\ \le\ \liminf_{n\to\infty}\frac{\min I}{n^3}\ \le\ \limsup_{n\to\infty}\frac{\min I}{n^3}\ \le\ \frac k3,
  \qquad
  c_k=\max_{m\ge 3}\ \frac{k}{m^k}\Bigl(\frac13-\frac1m\Bigr)>0 .
\]
\end{theorem}

\begin{proof}
Cut each axis of $[n]^k$ into $M$ intervals of at most $s$ consecutive values. This
cuts the grid into $M^k$ boxes. Two points in the same box differ by at most $s-1$ in every
coordinate, so their distance is at most $k(s-1)\le t$. If box $b$ holds $m_b$ of the $n$ points,
convexity gives
\[
  \sum_b\binom{m_b}{2}\ \ge\ M^k\binom{n/M^k}{2}=\frac n2\Bigl(\frac n{M^k}-1\Bigr)
\]
pairs that share a box. Each of them has $t-d\ge t-k(s-1)\ge0$. Inserting this
into~\eqref{eq:half} proves the first claim. For the second, take $s=\ceil{n/m}$ for a
fixed $m\ge3$. Then $M\le m$ and $s-1\le(n-1)/m$, and the bound becomes
$I\ge k(\frac13-\frac1m)(n^3/m^k-n^2)$ (\leanref{lower_bound_explicit}), which is
$k\,m^{-k}(\frac13-\frac1m)\,n^3-O(n^2)$ (\leanref{eventually_lower_bound}). For the upper bound
(\leanref{imbalance_le}), $k\le d\le k(n-1)$ gives
$\abs{d-t}\le\frac{2k(n-2)}3$, and $\binom n2\cdot\frac{2kn}{3}\le\frac{k n^3}{3}$.
\end{proof}

\noindent The Lean proof (\cref{sec:lean}) follows these lines and does the convexity step with the
Cauchy--Schwarz inequality $(\sum_b m_b)^2\le M^k\sum_b m_b^2$.

The upper constant $k/3$ is crude. Random rectangles do much better, though we have not
formalised this.

\begin{proposition}\label{prop:random}
Let $U_1,U_1',\dots,U_k,U_k'$ be independent and uniform on $[0,1]$, put
$Z_k=\sum_{i=1}^k\abs{U_i-U_i'}$ and $D_k=\frac12\,\mathbb E\abs{Z_k-\frac k3}$. Then for every fixed
$k$, $\min I(k\times n)\le D_k\,n^3+O(n^2)$, and
\[
  D_k\ \le\ \tfrac12\sqrt{\operatorname{Var}Z_k}\ =\ \frac{\sqrt k}{6\sqrt2}.
\]
In particular $\limsup_n\min I/n^3\le D_k$ in \cref{thm:cubic}.
\end{proposition}

\begin{proof}
Choose $r_1,\dots,r_k$ independently and uniformly from $\Sym_n$, ignoring the column condition.
For a fixed pair $u\ne v$ the pairs $(c_i(u),c_i(v))$ are independent across $i$ and uniform over
ordered pairs of distinct columns. Realise them as $\ceil{nU_i}$ and $\ceil{nU_i'}$ conditioned on
being distinct, an event of probability $1-1/n$. Rounding changes each coordinate difference by less than $1$. Conditioning on the $k$ events changes
the joint law by $O(k/n)$ in total variation, while $\abs{d-t}=O(kn)$. Together with
$t-\frac{kn}3=\frac k3$, this gives $\mathbb E\abs{d(u,v)-t}=n\,\mathbb E\abs{Z_k-\frac k3}+O(k^2)$. Summing over the $\binom n2$ pairs, $\mathbb E\,I=D_kn^3+O(n^2)$.
The number $Q$ of clashes, that is, of triples $(i,i',j)$ with $i<i'$ and the same symbol in
column $j$ of rows $i$ and $i'$, has $\mathbb E Q=\sum_{i<i'}\mathbb E\,\mathrm{fix}(r_i^{-1}r_{i'})=\binom k2$.
Choose an outcome with $I+4kn\,Q\le\mathbb E[I+4kn\,Q]=D_kn^3+O(n^2)$.

Now repair the clashes one at a time. Suppose symbol $a$ sits in column $j$ of row $i$ and also of another row. Swap it with the symbol $b$
in a column $j'$ of row $i$ with $1\le\abs{j-j'}\le2k$, chosen so that, in the other rows, $a$ does not
occur in column $j'$ and $b$ does not occur in column $j$.
At most $2k-2$ columns are excluded, so such a $j'$ exists once $n\ge4k$. The swap removes the clash
of $a$, creates none, and changes at most $2(n-2)$ distances by at most $2k$ each, so it changes $I$
by at most $4kn$. After at most $Q$ swaps the rectangle is Latin, with $I\le D_kn^3+O(n^2)$. Finally,
$\mathbb E Z_k=\frac k3$ and $\operatorname{Var}\abs{U-U'}=\frac16-\frac19=\frac1{18}$, so the
Cauchy--Schwarz inequality gives $\mathbb E\abs{Z_k-\frac k3}\le\sqrt{k/18}$.
\end{proof}

The proof of \cref{thm:cubic} is a pigeonhole argument, and \cref{fig:geometry}a shows it at work.
Clustered points are expensive, and in a low-dimensional grid clustering is unavoidable. Three
remarks put the theorem in context.

\begin{remark}[Constants]\label{rem:constants}
For $k=2$, the maximum that defines $c_2$ is attained at $m=5$, which gives
$c_2=4/375\approx0.0107$ (\leanref{lowerConst_two_five}, \leanref{lowerConst_two_le}). Also
$D_2=496/3645\approx0.1361$, the value for a uniformly random derangement. For $k\ge3$,
$c_k=k/(12\cdot4^k)$, attained at $m=4$ (\leanref{lowerConst_four}, \leanref{lowerConst_le_four}).
Numerically, $D_k\approx0.094\sqrt k$ for large $k$, for example $D_3\approx0.165$ and
$D_{12}\approx0.327$. So the lower constant is of order $k\,4^{-k}$, and the upper one grows like
$\sqrt k$. We do not know the right order in $k$.

In the benchmark range, the finite-$n$ lower bound is weak. For $2\times13$ it gives $7.7$, against
the true optimum $218.6$, and for $k=3$ it becomes positive only at $n=28$. Its message is
asymptotic. For fixed $k$, the average deviation per pair, $I/\binom n2$, grows linearly in $n$, the
same order as the target $t$ itself. So rectangles are imperfect on average at the scale of $t$.
\end{remark}

\begin{remark}[Squares]
The bound is useless when $k$ grows with $n$, and it must be. Spatially balanced Latin squares exist,
for instance whenever $2n+1$ is prime, by a construction that streamlined search itself
discovered~\citep{LeBrasGomesSelman2012}. In $[n]^n$ there is room for
$n$ equidistant points. For squares with $n\equiv1\pmod3$, where perfect balance is impossible,
\citet{XiaEtAl2026} prove $I\ge4n(n-1)/9$, an order $n^2$ rather than $n^3$. The bound is attained for $n\equiv1\pmod 3$ up to $n=52$~\citep{XiaEtAl2026}. It is not known
whether the minimum for squares is of order at most $n^2$ for every $n$. The point picture raises a
natural question, which we leave open: how does the minimum imbalance pass from order $n^3$ at
fixed $k$ to these much smaller values at $k=n$?
\end{remark}

\begin{remark}[Equilateral sets]\label{rem:equilateral}
A perfectly balanced rectangle needs $3\mid k(n+1)$, so that $t$ is an integer. It is also a set of
$n$ points in $\R^k$ that are pairwise equidistant in $\ell_1$, that is, an \emph{equilateral set} of
$\ell_1^k$. Hence $n\le e(\ell_1^k)$, the largest size of such a set. Kusner conjectured $e(\ell_1^k)=2k$. The conjecture is known for $k\le4$, where the case $k=4$ is due
to \citet{KoolenLaurentSchrijver2000}. In general, \citet{AlonPudlak2003} proved
$e(\ell_1^k)=O(k\log k)$. The survey of \citet{Swanepoel2004} covers these results. So a
perfectly balanced $k\times n$ rectangle needs $k=\Omega(n/\log n)$ rows. For $k\le4$ this gives
$n\le2k$, which recovers part of the case analysis of~\citet{DiazLeBrasGomes2017}. Their arguments are sharper for $k\le 3$. Their bound $k\ge 6-18/(n+1)$ coincides with ours at $k=4$
and is vacuous for $k\ge6$. The point sets of rectangles are far more special than general
equilateral sets, because their coordinate projections are bijections onto $[n]$. So we do not
expect these bounds to be tight. We did not find this connection in the literature on balanced
designs.
\end{remark}

\FloatBarrier
\section{Two rows}\label{sec:two}

For $k=2$ a rectangle is a derangement $\pi=r_2$, and the points $x_u=(u,\pi^{-1}(u))$ form its permutation diagram. \citet{MirkaEtAl2023}
solved $n\le13$ to optimality with a clique-based MIP and reported bounds up to $n=20$ after six
hours. \Cref{tab:two} adds our two-minute annealing runs. They improve the upper bound for $15\le n\le 19$, tie at $n=14$ and lose at $n=20$. At $n=20$, the
best bound of~\citep{MirkaEtAl2023} came from eight parallel six-hour runs. Our runs are single runs
with one seed, and the improvements are small ($0.2$--$2\%$). The MIP also spends its time on lower
bounds. So the comparison shows only that annealing is a strong source of incumbents. The lower
bounds remain far away, and we did not try to improve them.

\begin{table}[t]
\centering
\small
\input{table_k2.tex}
\caption{$2\times n$ rectangles beyond the proven range. The lower and upper bounds are the best
over all methods and configurations in Tables 1, 2 and 4--7 of~\citep{MirkaEtAl2023}, from 6-hour runs. Annealing:
one 2-minute single-core run. Bold: better than the previous upper bound. The lower bounds are
those reached when the 6-hour limit expired, which is why they are not monotone in $n$. The values $201.57$ and $273.82$ come from one instance
with 32 threads in their Table~6 and are decimals as reported there. All other values use the $x.3$,
$x.6$ convention.}
\label{tab:two}
\end{table}

\begin{figure}[t]
\centering
\includegraphics[width=\linewidth]{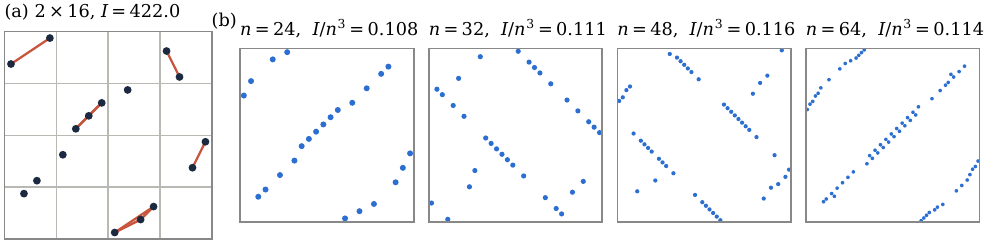}
\caption{$2\times n$ rectangles as point sets. (a)~The proof of \cref{thm:cubic} for a
$2\times16$ rectangle with $I=422$: a $4\times4$ grid of boxes, and in red the pairs that share
a box. Each such pair is closer than the average $t=34/3$ and costs imbalance.
(b)~The best rectangles found by annealing for larger $n$ (one 2-minute run each), rescaled to
the unit square. Most points lie on a few long parallel diagonal runs, with scattered extra
points.}
\label{fig:geometry}
\end{figure}

\begin{figure}[t]
\centering
\includegraphics[width=0.9\linewidth]{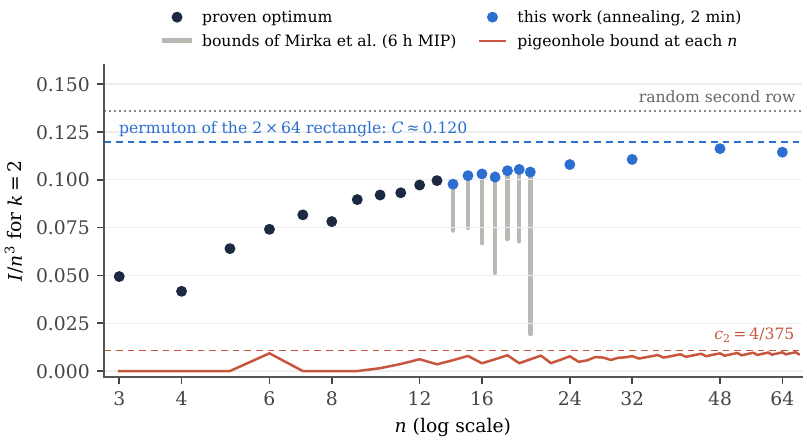}
\caption{Normalised imbalance $I/n^3$ for $k=2$. Black: proven optima. Grey: interval between
the best lower and upper bounds of~\citep{MirkaEtAl2023}. Blue dots: our annealing values. Blue dashes: $C(\mu_\sigma)$ for our $2\times64$ rectangle,
an upper bound for the limit (\cref{prop:limit}). Red:
\cref{thm:cubic} at each $n$ (solid) and its limit $c_2$ (dashed). A uniformly random
derangement has $I/n^3\to0.1361$ (dotted).}
\label{fig:cubic}
\end{figure}

\paragraph{The shape of good permutations.} \Cref{fig:geometry}b shows the best permutations
that annealing finds for $n\in\{24,32,48,64\}$. They are not known to be optimal, but they are far from random. Most points lie on a few
long parallel segments of slope $1$, cyclically shifted against one another, with a minority of
scattered points.

\paragraph{A limit object.} A \emph{permuton} is a probability measure on $[0,1]^2$ with
uniform marginals. Every permutation $\sigma$ of $[n]$ defines one, $\mu_\sigma$, uniform on
the squares $[\frac{i-1}n,\frac in]\times[\frac{\sigma(i)-1}n,\frac{\sigma(i)}n]$. Put
\[
  C(\mu)=\tfrac12\,\mathbb E\,\bigl|\,\abs{X-X'}+\abs{Y-Y'}-\tfrac23\,\bigr|,
\]
where $(X,Y)$ and $(X',Y')$ are independent samples of $\mu$.

\begin{proposition}\label{prop:limit}
For $k=2$, $\min I/n^3$ converges as $n\to\infty$ to $\min_\mu C(\mu)$, where the minimum over
permutons is attained.
\end{proposition}

\begin{proof}
Let $f(x,y,x',y')=\bigl|\abs{x-x'}+\abs{y-y'}-\frac23\bigr|$, which is bounded and $1$-Lipschitz
for the $\ell_1$ norm on $[0,1]^4$. If $\mu_n\to\mu$ weakly then $\mu_n\otimes\mu_n\to\mu\otimes\mu$ weakly, so $C$ is
continuous. Permutons form a weakly closed subset of the compact space of probability measures on
$[0,1]^2$, so $C$ attains its minimum.

Fix a derangement $\sigma$ of $[n]$ and take the $2\times n$ rectangle with $\pi=\sigma^{-1}$,
whose points $x_u=(u,\sigma(u))$ are the cells of $\mu_\sigma$. Write
$I(\sigma)/n^3=\frac1{2n^2}\sum_{u\ne v}\abs{d(u,v)/n-t/n}$. Each ordered pair $u\ne v$ corresponds to
a pair of cells of $\mu_\sigma\otimes\mu_\sigma$ of weight $1/n^2$. Within these cells, $f$ differs
from $\abs{d(u,v)/n-\frac23}$ by at most $2/n$, and $\abs{t/n-\frac23}=\frac2{3n}$. The $n$
diagonal cell pairs carry total weight $1/n$ and are absent from $I$. Hence
$\abs{I(\sigma)/n^3-C(\mu_\sigma)}\le A/n$ for an absolute constant $A$, uniformly in $\sigma$.
This gives $\min I/n^3\ge\min_\mu C(\mu)-A/n$.

Conversely, let $\mu^\ast$ attain the minimum. Sampling $n$ independent points from $\mu^\ast$ and
taking the induced permutation gives permutations $\sigma_n$ with $\mu_{\sigma_n}\to\mu^\ast$
almost surely~\citep{HoppenEtAl2013}. Remove the fixed points of $\sigma_n$ one at a time: if
$\sigma(j)=j$, exchange the values at $j$ and $j'$, where $j'=j+1$ for $j<n$ and $j'=n-1$ for
$j=n$. Then $\sigma(j)$ becomes $\sigma(j')\ne j$ and $\sigma(j')$ becomes $j\ne j'$, so no new
fixed point is created. Each exchange moves two points by one cell, so it changes $I$ by at most
$4n$, and there are at most $n$ of them. The resulting derangements $\tau_n$ satisfy
$I(\tau_n)/n^3\le C(\mu_{\sigma_n})+O(1/n)\to C(\mu^\ast)$.
\end{proof}

The uniform permuton gives $C=496/3645\approx0.1361$, the limit for a random derangement, and the
diagonal (the cyclic rectangle) gives $16/81\approx0.198$. Optimising numerically over permutons
made of $m$ increasing or decreasing segments gives $0.148$, $0.133$ and $0.124$ for $m=3,4,5$.
The permuton of our $2\times64$ rectangle does better. On a pair of distinct cells,
$\abs{X-X'}+\abs{Y-Y'}$ is a constant plus a sum of four independent uniform variables. So $C(\mu_\sigma)$ is a finite sum
of explicit polynomial integrals, and exact rational arithmetic gives $C(\mu_\sigma)=228509453/1911029760\approx0.119574$. This is our best upper bound for
the limit. The rectangle itself has $I=29988$, so $I/n^3=0.11440$. The diagonal cells account for almost all of the difference $0.00518$. They contribute $\frac1{3n}-\frac1{3n^2}=0.00513$ to $C(\mu_\sigma)$,
and the remaining $0.00005$ comes from the within-cell smoothing and from $t/n\ne\frac23$. The finite
values $I/n^3\approx0.11$ in \cref{fig:cubic} therefore lie below the limit estimate by about
$1/(3n)$.

\begin{question}
What is $\lim_n\min I(2\times n)/n^3=\min_\mu C(\mu)$? We know only that it lies between
$4/375\approx0.011$ and $0.1196$. Is the minimiser unique, and is it a finite union of
segments?
\end{question}

\FloatBarrier
\section{Formal verification in Lean 4}\label{sec:lean}

Most computational results on balanced rectangles rest on unverified programs. The DRAT proofs
of~\citet{RamaswamySzeider2023} are an exception, and for squares \citet{XiaEtAl2026} verified
their lower bound in Lean~4. We formalised the parts of
this paper where formalisation is cheap and useful, in Lean~4~\citep{deMouraUllrich2021} with
Mathlib~\citep{mathlib2020}. The development has about 700 hand-written lines plus generated
certificates. It is in the \texttt{lean/} directory of the repository, and
\cref{tab:lean} maps the statements of the paper to Lean names.

\begin{table}[t]
\centering
\small
\setlength{\tabcolsep}{4pt}
\begin{tabular}{@{}lll@{}}
\toprule
statement in the paper & Lean name & uses \\
\midrule
all rectangles (\cref{tab:results,tab:two}, \cref{fig:optimum}) & \leanref{Certificates.exists_k*_n*} & core \\
$I=\mathrm{imb3}/3$; $t$ is the mean distance & \leanref{imbalance_eq_imb3}, \leanref{sum_dist} & Mathlib \\
identity~\eqref{eq:half} & \leanref{imbalance_eq_two_mul_sum_pos} & Mathlib \\
\cref{thm:cubic}, pigeonhole bound & \leanref{cubic_lower_bound_ceil} & Mathlib \\
\cref{thm:cubic}, $\liminf\ge c_k$ & \leanref{eventually_lower_bound} & Mathlib \\
\cref{thm:cubic}, $\limsup\le k/3$ & \leanref{imbalance_le} & Mathlib \\
$c_2=4/375$ & \leanref{lowerConst_two_five}, \leanref{lowerConst_two_le} & Mathlib \\
$c_k=k/(12\cdot4^k)$, $k\ge3$ & \leanref{lowerConst_four}, \leanref{lowerConst_le_four} & Mathlib \\
\midrule
rectangles of~\citep{RamaswamySzeider2023,VoborilRamaswamySzeider2025} & \leanref{Certificates.exists_pub_*} & core \\
spec = Latin arrays; non-vacuous & \leanref{arrayEquiv}, \leanref{nonempty_of_le} & Mathlib \\
\midrule
optimality in \cref{thm:3x9} & \multicolumn{2}{l}{not formalised} \\
\cref{prop:rowgroup,prop:random,prop:limit} & \multicolumn{2}{l}{not formalised} \\
\bottomrule
\end{tabular}
\caption{Formalised statements. Names are relative to the namespace \leanref{BalancedLatin};
theorems about a rectangle are in its sub-namespace \leanref{LatinRect}.
``Core'': the proof uses only Lean's core library and the 74-line specification
\texttt{Defs.lean}; ``Mathlib'': it also uses Mathlib. The certificates use only the axiom \leanref{propext}; the
theorems use the three standard axioms \leanref{propext}, \leanref{Classical.choice} and
\leanref{Quot.sound}. No statement uses \leanref{sorry} or \leanref{native_decide}.}
\label{tab:lean}
\end{table}

\paragraph{Specification.} The file \texttt{Defs.lean} contains the whole specification that the
certificates rely on. It imports only Lean's core library, not Mathlib. It uses the point picture
of \cref{sec:prelim}: a rectangle is the function $(i,u)\mapsto c_i(u)$, with indices starting
at~$0$.
\par\noindent\begin{minipage}{\linewidth}
\begin{lstlisting}
structure LatinRect (k n : Nat) where
  col : Fin k $\to$ Fin n $\to$ Fin n                -- col i u = column of symbol u in row i
  row_inj : $\forall$ i u v, col i u = col i v $\to$ u = v
  col_inj : $\forall$ u i j, col i u = col j u $\to$ i = j

def dist (L : LatinRect k n) (u v : Fin n) : Nat :=     -- $d(u,v)=\sum_i|c_i(u)-c_i(v)|$
  fsum fun i => absDiff (L.col i u).val (L.col i v).val

def imb3 (L : LatinRect k n) : Nat :=                   -- $3I=\sum_{u<v}|3d(u,v)-k(n+1)|$
  fsum fun u => fsum fun v =>
    if u < v then ((3 * L.dist u v : Int) - k * (n + 1)).natAbs else 0
\end{lstlisting}
\end{minipage}
Here \leanref{fsum f} is the sum of the list \leanref{[f 0, ..., f (n-1)]}, and
\leanref{absDiff a b} is $(a\mathbin{\dot-}b)+(b\mathbin{\dot-}a)=\abs{a-b}$, with truncated
subtraction~$\dot-$ on $\N$. We certify the integer $3I$, so
that the kernel never handles fractions. In \texttt{Cubic.lean}, we define the imbalance $I$ over $\mathbb Q$ exactly as in \cref{sec:prelim},
and \leanref{imbalance_eq_imb3} proves $I=\mathrm{imb3}/3$.

\paragraph{Certificates by kernel evaluation.} A script copies each rectangle from its data file
into Lean, after cross-checking the claimed value with the independent Python checker. For
\cref{fig:optimum} it produces
\begin{lstlisting}
def rows_k3_n9 : List (List Nat) :=
  [[1, 2, 3, 4, 5, 6, 7, 8, 9], [8, 3, 1, 6, 7, 4, 5, 9, 2], [4, 5, 8, 1, 9, 3, 2, 6, 7]]

def rect_k3_n9 : LatinRect 3 9 where
  col := colOfRows rows_k3_n9
  row_inj := by decide +kernel
  col_inj := by decide +kernel

theorem imb3_k3_n9 : rect_k3_n9.imb3 = 168 := by decide +kernel
theorem exists_k3_n9 : $\exists$ L : LatinRect 3 9, L.imb3 = 168 := $\langle$rect_k3_n9, imb3_k3_n9$\rangle$
\end{lstlisting}
The tactic \leanref{decide +kernel} is proof by reflection. The proposition has a decision
procedure, written in Lean, and the kernel itself reduces that procedure to \leanref{true} while
type-checking the proof term. The alternative, \leanref{native_decide}, runs compiled code. It
adds the axiom \leanref{Lean.ofReduceBool}, so it trusts the compiler and every
\leanref{implemented_by} annotation. Our certificates avoid it, and \texttt{\#print axioms}
reports only \leanref{propext}.

The soundness of a statement such as \leanref{exists_k3_n9} does not depend on the script or on
how \leanref{colOfRows} reads the data. For any function passed as \leanref{col}, the kernel checks that it is a Latin rectangle with the
stated value of $3I$. To tie each certificate
to its data file, a further theorem \leanref{faithful_k3_n9} checks that the entry of row~$i$ in
column \leanref{col i u} is the symbol $u+1$ of the file. Lean certifies all 62 rectangles: the 51 of \cref{tab:results} and the eleven $2\times n$ rectangles with $14\le n\le64$ used
in \cref{tab:two,fig:geometry,fig:cubic}. The largest are $12\times13$ and $2\times64$, which has
2016 pairs. Elaborating all of them takes about 50\,s of CPU time on an Apple M3 laptop. As a
negative test, changing the claimed value by one, placing a symbol twice in a column, or
repeating a symbol in a row each makes the check fail.

\paragraph{Cubic growth.} \texttt{Cubic.lean} (480 lines, 14\,s) proves \cref{thm:cubic}. The
main statements read
\par\noindent\begin{minipage}{\linewidth}
\begin{lstlisting}
theorem cubic_lower_bound (L : LatinRect k n) (s M : $\mathbb N$) (hs : 0 < s) (hM : n $\le$ M * s)
    (ht : (k : $\mathbb Q$) * (s - 1) $\le$ target k n) :
    (n : $\mathbb Q$) * (n / (M : $\mathbb Q$) ^ k - 1) * (target k n - k * (s - 1)) $\le$ L.imbalance

theorem eventually_lower_bound (k m : $\mathbb N$) (hm : 3 $\le$ m) {$\varepsilon$ : $\mathbb Q$} (h$\varepsilon$ : 0 < $\varepsilon$) :
    $\forall^{\mathsf f}$ n in Filter.atTop, $\forall$ L : LatinRect k n,
      (lowerConst k m - $\varepsilon$) * (n : $\mathbb Q$) ^ 3 $\le$ L.imbalance
\end{lstlisting}
\end{minipage}
where \leanref{target k n} is $t=k(n+1)/3$ and \leanref{lowerConst k m} is
$k\,m^{-k}(\frac13-\frac1m)$. The first is slightly more general than \cref{thm:cubic}: it
allows any $M$ intervals per axis with $n\le Ms$. Its instance $M=\ceil{n/s}$,
\leanref{cubic_lower_bound_ceil}, is \cref{thm:cubic} word for word. The second states the liminf bound in the
eventually form, for each $m$, so we do not have to define $\min I$ as a function of~$n$. The
proof follows the paper. It reindexes along the row permutations (\leanref{Fintype.sum_bijective}) to show that $t$ is the
average distance. It then symmetrises the sum over $u<v$ to one over ordered pairs, and uses Mathlib's \leanref{sq_sum_le_card_mul_sum_sq} for
Cauchy--Schwarz.

\paragraph{Is the specification the right one?} A formal proof is only as good as its statement. A definition that differs subtly from the intended
one would make every certificate meaningless. We guard against this in four ways.
(i)~\emph{Agreement with the literature.} Lean certifies the 14 rectangles printed
in~\citep{RamaswamySzeider2023,VoborilRamaswamySzeider2025}, of sizes $3\times4$ to $12\times13$, with
the imbalances that those papers report
(\leanref{Certificates.exists_pub_*}). This tests the definitions of $d$ and $I$ against
independent data, not against our own Python checker.
(ii)~\emph{Agreement with the textbook definition.} \leanref{arrayEquiv} is a bijection between
\leanref{LatinRect k n} and arrays $A$ (row $i$, column $j$) whose rows are permutations and whose columns have no repeats. \leanref{ofArray_col} shows that \leanref{col i u} is the column of
symbol $u$ in row $i$, that is, $c_i(u)$.
(iii)~\emph{Non-vacuity.} \leanref{nonempty_of_le} shows that \leanref{LatinRect k n} is
inhabited for every $k\le n$, so the universally quantified theorems are not vacuous.
(iv)~\emph{Agreement with the paper.} A test parses \cref{tab:results,tab:two} and the
values quoted in the text for $3\times9$, $2\times16$ and $12\times13$, and compares them with the
Lean statements. It also compares every rectangle printed in \cref{app:rectangles}, entry by entry,
with the rows of its certificate. Changing one of these printed values makes it fail.

\paragraph{What is not formalised.} The optimality half of \cref{thm:3x9} is the main gap. The
branch and bound visits $5.5\cdot10^9$ nodes, far more than kernel evaluation can handle. A certificate would need one of two routes. The first is a branch and bound written and proved
correct in Lean and then run with \leanref{native_decide}, which moves the trust from our C program to
Lean's compiler. The second is a SAT encoding with an LRAT proof that a verified checker accepts. Either route also has to prove the
symmetry reduction of \cref{sec:exact}. \Cref{prop:limit} needs permutons and the sampling theorem
of~\citet{HoppenEtAl2013}, which Mathlib does not have. \Cref{prop:random} needs an asymptotic
estimate of $\mathbb E\abs{Z_k-\frac k3}$ for discretised uniform variables. \Cref{rem:equilateral} rests on the
theorem of \citet{AlonPudlak2003}. \Cref{prop:rowgroup} is elementary, but we computed the row groups in \cref{fig:rowgroups} with
SymPy's Schreier--Sims implementation, and they are not certified.

\section{Discussion}

The point picture says that balance is a packing problem in $\ell_1$, so moving points is the natural
search move. It also gives the right scale. For a fixed number of rows, imbalance is not a small
defect that clever search will remove. It is a quantity of order $n^3$, and its leading constant is
the real object of study. The row group adds a cheap check that a rectangle is not built from a
group table.

\paragraph{Limitations.} The new upper bounds are best-of-runs from at most three runs per
instance, and single runs vary by up to a third (\cref{sec:results}). We did not isolate the effect of the adaptive penalty or rerun earlier local search head-to-head. We
also make no runtime comparison with earlier work. The optimality of $3\times 9$ rests on two
uncertified programs that share a branching scheme. In contrast, Lean checks every upper bound and
the pigeonhole theorem.
Our lower-bound constant is far from the observed values for $k=2$ and is negligible in the
benchmark range.

\paragraph{Open problems.}
\begin{enumerate}[label=(\arabic*),leftmargin=*,itemsep=2pt]
\item Certify \cref{thm:3x9}, for instance with a DRAT-producing SAT encoding as
in~\citep{RamaswamySzeider2023} or with a branch and bound verified in Lean (\cref{sec:lean}). Also
settle $5\times8$ to $7\times8$. The sharper bound and the unused column-reversal and row symmetries
of \cref{sec:exact} are natural first steps. Our new upper bounds are natural starting incumbents for
certified solvers. The mirror pairing $u\leftrightarrow n+1-u$ seen in several records is a candidate
streamliner in the sense of~\citep{VoborilRamaswamySzeider2025}.
\item Close the gap between $c_2=0.0107$ and $0.1196$. A sharper lower bound might weight each close
pair by its shortfall $t-d$ instead of the worst case inside a box. The limit problem is a quadratic
minimisation over permutons. One route to a certified lower bound is to discretise it to doubly
stochastic matrices and apply a semidefinite relaxation.
\item Determine the minimising permuton for $k=2$.
\item Find the order in $k$ of the constants in \cref{thm:cubic}, between $k\,4^{-k}$ and $\sqrt k$.
Also find how the minimum passes to order $n^2$ for squares.
\item Show directly that a perfectly balanced $k\times n$ rectangle needs $k\ge n/2$, using that its
coordinate projections are bijections. This would follow from Kusner's conjecture for $\ell_1$, a
long-standing problem in geometry, but it may be easier for these special point sets.
\end{enumerate}

\paragraph{Data and code.} Rectangles, the annealing code, the branch and bound, the exact
checker, the Lean development and the scripts for every figure are at
\\\url{https://github.com/BrianLi009/Balanced-Latin-Rectangles}. The new rectangles are listed in \cref{app:rectangles}, and all 62 are in the ancillary file \texttt{anc/rectangles.txt}.

\FloatBarrier
\bibliographystyle{plainnat}
\bibliography{refs}

\clearpage
\appendix
\section{The new rectangles}\label{app:rectangles}

This appendix lists the \NumImproved{} rectangles that improve \cref{tab:results} and the seven
$2\times n$ rectangles of \cref{tab:two}. Rows are listed from top to bottom, entry $j$ of a row is
the symbol in column $j$, and the first row is always the identity. Each heading gives the new
imbalance in bold and, in grey, the previous best value of~\citep{VoborilRamaswamySzeider2025} and the
relative improvement. As in \cref{fig:optimum}, each symbol has its own colour, so every row shows
all $n$ colours and no colour repeats within a column.
Each rectangle is a file of the repository named by its size: the $4\times9$ rectangle, for
example, is \texttt{results/rectangles/k4\_n9.txt} (the $2\times n$ ones are in
\texttt{results/k2\_large/}). Lean certifies it as \leanref{Certificates.exists_k4_n9}, and \leanref{Certificates.faithful_k4_n9}
ties the certificate to the file (\cref{sec:lean}). A test in the repository checks this list against
the certificates entry by entry. All 62 rectangles of the paper, including the matched records and the $2\times n$
rectangles with $n\ge24$, are in the ancillary file \texttt{anc/rectangles.txt}.

\input{appendix_rects.tex}

\end{document}

%% file: numbers.tex
\newcommand{\NumCells}{51}
\newcommand{\NumImproved}{33}
\newcommand{\NumMatched}{18}

\newcommand{\MeanGain}{18.5}
\newcommand{\MaxGain}{39.4}
\newcommand{\TopCell}{$12\times13$}
\newcommand{\TopOld}{274.0}
\newcommand{\TopNew}{166.0}

\newcommand{\NumTimingRuns}{165}
\newcommand{\NumSeeds}{5}

\newcommand{\MedianBeat}{0.12}
\newcommand{\MaxBeat}{143}

%% file: table.tex
\begin{tabular}{rccccccccccc}
\toprule
$n \backslash k$ & 2 & 3 & 4 & 5 & 6 & 7 & 8 & 9 & 10 & 11 & 12 \\
\midrule
8 & 40.0 & 36.0 & 32.0 & 30.0 & 24.0 & 28.0 &  &  &  &  &  \\[2pt]
9 & 65.3 & 56.0 & \shortstack{\textbf{53.3}\\{\tiny\sout{56.0}}} & \shortstack{\textbf{48.6}\\{\tiny\sout{52.6}}} & \shortstack{\textbf{40.0}\\{\tiny\sout{48.0}}} & \shortstack{\textbf{43.3}\\{\tiny\sout{51.3}}} & \shortstack{\textbf{42.6}\\{\tiny\sout{53.3}}} &  &  &  &  \\[2pt]
10 & 92.0 & 80.0 & 78.6 & 66.6 & \shortstack{\textbf{60.0}\\{\tiny\sout{70.0}}} & \shortstack{\textbf{58.0}\\{\tiny\sout{72.0}}} & \shortstack{\textbf{58.6}\\{\tiny\sout{82.6}}} & \shortstack{\textbf{64.0}\\{\tiny\sout{70.0}}} &  &  &  \\[2pt]
11 & 124.0 & 114.0 & \shortstack{\textbf{104.0}\\{\tiny\sout{112.0}}} & \shortstack{\textbf{94.0}\\{\tiny\sout{116.0}}} & \shortstack{\textbf{88.0}\\{\tiny\sout{108.0}}} & \shortstack{\textbf{84.0}\\{\tiny\sout{118.0}}} & \shortstack{\textbf{78.0}\\{\tiny\sout{114.0}}} & \shortstack{\textbf{84.0}\\{\tiny\sout{120.0}}} & \shortstack{\textbf{100.0}\\{\tiny\sout{110.0}}} &  &  \\[2pt]
12 & 168.0 & \shortstack{\textbf{152.0}\\{\tiny\sout{154.0}}} & \shortstack{\textbf{149.3}\\{\tiny\sout{154.6}}} & \shortstack{\textbf{140.0}\\{\tiny\sout{145.3}}} & 120.0 & \shortstack{\textbf{121.3}\\{\tiny\sout{164.0}}} & \shortstack{\textbf{116.6}\\{\tiny\sout{148.0}}} & \shortstack{\textbf{106.0}\\{\tiny\sout{154.0}}} & \shortstack{\textbf{124.0}\\{\tiny\sout{174.6}}} & \shortstack{\textbf{136.6}\\{\tiny\sout{147.3}}} &  \\[2pt]
13 & 218.6 & 198.0 & \shortstack{\textbf{201.3}\\{\tiny\sout{211.3}}} & \shortstack{\textbf{191.3}\\{\tiny\sout{210.6}}} & \shortstack{\textbf{166.0}\\{\tiny\sout{204.0}}} & \shortstack{\textbf{164.0}\\{\tiny\sout{203.3}}} & \shortstack{\textbf{160.0}\\{\tiny\sout{220.6}}} & \shortstack{\textbf{152.0}\\{\tiny\sout{214.0}}} & \shortstack{\textbf{154.6}\\{\tiny\sout{240.6}}} & \shortstack{\textbf{168.0}\\{\tiny\sout{224.0}}} & \shortstack{\textbf{166.0}\\{\tiny\sout{274.0}}} \\[2pt]
\bottomrule
\end{tabular}

%% file: table_k2.tex
\begin{tabular}{@{}lrrrrrrr@{}}
\toprule
$n$ & 14 & 15 & 16 & 17 & 18 & 19 & 20 \\
\midrule
lower bound \citep{MirkaEtAl2023} & 201.57 & 252.6 & 273.82 & 252.0 & 402.6 & 465.3 & 155.0 \\
upper bound \citep{MirkaEtAl2023} & 268.0 & 345.3 & 427.3 & 508.0 & 617.3 & 732.0 & 816.0 \\
annealing (this work)             & 268.0 & \textbf{344.6} & \textbf{422.0} & \textbf{498.0} & \textbf{610.6} & \textbf{722.6} & 832.0 \\
\bottomrule
\end{tabular}

%% file: appendix_rects.tex
\begingroup
\definecolor{blrdark}{HTML}{1B2A41}
\newcommand*{\blc}[4]{\definecolor{blr#1-#2}{HTML}{#3}\colorlet{blrt#1-#2}{#4}}
\newlength{\blrslot}
\newcommand{\blrcard}[3]{%
  \begin{minipage}[t]{\blrslot}\centering
  {\scriptsize #1}\par\vspace{3pt}
  \begin{tikzpicture}[x=3.8mm,y=-3.8mm]
  \foreach \row [count=\i] in {#3} {\foreach \s [count=\j] in \row {
    \fill[blr#2-\s] (\j-1,\i-1) rectangle ++(0.92,0.92);
    \node[text=blrt#2-\s,font=\fontsize{7}{8}\selectfont,inner sep=0pt] at (\j-0.54,\i-0.54) {\s};}}
  \end{tikzpicture}\end{minipage}}
\blc{9}{1}{4C176B}{white}\blc{9}{2}{5E4FAC}{white}\blc{9}{3}{6A8BBF}{white}\blc{9}{4}{A3BECA}{blrdark}\blc{9}{5}{E2D9E2}{blrdark}\blc{9}{6}{CFAF99}{blrdark}\blc{9}{7}{BD6E5A}{white}\blc{9}{8}{973450}{white}\blc{9}{9}{561546}{white}
\blc{10}{1}{4C176B}{white}\blc{10}{2}{5E48A8}{white}\blc{10}{3}{657FBD}{white}\blc{10}{4}{8AAEC5}{blrdark}\blc{10}{5}{CED3D9}{blrdark}\blc{10}{6}{DCCECB}{blrdark}\blc{10}{7}{CA9A7D}{blrdark}\blc{10}{8}{B75F55}{white}\blc{10}{9}{912F50}{white}\blc{10}{10}{561546}{white}
\blc{11}{1}{4C176B}{white}\blc{11}{2}{5D42A4}{white}\blc{11}{3}{6275BA}{white}\blc{11}{4}{7BA1C2}{blrdark}\blc{11}{5}{B3C6CE}{blrdark}\blc{11}{6}{E2D9E2}{blrdark}\blc{11}{7}{D4BCAC}{blrdark}\blc{11}{8}{C6896C}{blrdark}\blc{11}{9}{B15552}{white}\blc{11}{10}{8C2A50}{white}\blc{11}{11}{561546}{white}
\blc{12}{1}{4C176B}{white}\blc{12}{2}{5D3DA1}{white}\blc{12}{3}{606CB8}{white}\blc{12}{4}{7195C0}{blrdark}\blc{12}{5}{9CB9C8}{blrdark}\blc{12}{6}{D3D6DB}{blrdark}\blc{12}{7}{DDD1D1}{blrdark}\blc{12}{8}{CEA991}{blrdark}\blc{12}{9}{C27A61}{blrdark}\blc{12}{10}{AC4C50}{white}\blc{12}{11}{872750}{white}\blc{12}{12}{561546}{white}
\blc{13}{1}{4C176B}{white}\blc{13}{2}{5D3A9E}{white}\blc{13}{3}{5F65B5}{white}\blc{13}{4}{6A8BBF}{white}\blc{13}{5}{8AAEC5}{blrdark}\blc{13}{6}{BECCD2}{blrdark}\blc{13}{7}{E2D9E2}{blrdark}\blc{13}{8}{D7C3B8}{blrdark}\blc{13}{9}{CA9A7D}{blrdark}\blc{13}{10}{BD6E5A}{white}\blc{13}{11}{A74650}{white}\blc{13}{12}{842550}{white}\blc{13}{13}{561546}{white}
\blc{14}{1}{4C176B}{white}\blc{14}{2}{5C369B}{white}\blc{14}{3}{5F5EB3}{white}\blc{14}{4}{6683BD}{white}\blc{14}{5}{7EA4C3}{blrdark}\blc{14}{6}{A9C1CB}{blrdark}\blc{14}{7}{D6D7DD}{blrdark}\blc{14}{8}{DED3D4}{blrdark}\blc{14}{9}{D1B4A0}{blrdark}\blc{14}{10}{C78C6F}{blrdark}\blc{14}{11}{B96456}{white}\blc{14}{12}{A24050}{white}\blc{14}{13}{802350}{white}\blc{14}{14}{561546}{white}
\blc{15}{1}{4C176B}{white}\blc{15}{2}{5C3499}{white}\blc{15}{3}{5F59B1}{white}\blc{15}{4}{637BBC}{white}\blc{15}{5}{769BC1}{blrdark}\blc{15}{6}{97B7C7}{blrdark}\blc{15}{7}{C5CFD5}{blrdark}\blc{15}{8}{E2D9E2}{blrdark}\blc{15}{9}{D9C8C0}{blrdark}\blc{15}{10}{CDA68C}{blrdark}\blc{15}{11}{C48166}{blrdark}\blc{15}{12}{B55C54}{white}\blc{15}{13}{9F3C50}{white}\blc{15}{14}{7E2250}{white}\blc{15}{15}{561546}{white}
\blc{16}{1}{4C176B}{white}\blc{16}{2}{5B3196}{white}\blc{16}{3}{5E53AE}{white}\blc{16}{4}{6275BA}{white}\blc{16}{5}{6F93C0}{blrdark}\blc{16}{6}{8AAEC5}{blrdark}\blc{16}{7}{B3C6CE}{blrdark}\blc{16}{8}{D8D8DE}{blrdark}\blc{16}{9}{DFD4D7}{blrdark}\blc{16}{10}{D4BCAC}{blrdark}\blc{16}{11}{CA9A7D}{blrdark}\blc{16}{12}{C1775F}{white}\blc{16}{13}{B15552}{white}\blc{16}{14}{9A3750}{white}\blc{16}{15}{7B2150}{white}\blc{16}{16}{561546}{white}
\blc{17}{1}{4C176B}{white}\blc{17}{2}{5B2F94}{white}\blc{17}{3}{5E4FAC}{white}\blc{17}{4}{616FB9}{white}\blc{17}{5}{6A8BBF}{white}\blc{17}{6}{80A5C3}{blrdark}\blc{17}{7}{A3BECA}{blrdark}\blc{17}{8}{CBD2D7}{blrdark}\blc{17}{9}{E2D9E2}{blrdark}\blc{17}{10}{DBCBC6}{blrdark}\blc{17}{11}{CFAF99}{blrdark}\blc{17}{12}{C78F72}{blrdark}\blc{17}{13}{BD6E5A}{white}\blc{17}{14}{AE5051}{white}\blc{17}{15}{973450}{white}\blc{17}{16}{792050}{white}\blc{17}{17}{561546}{white}
\blc{18}{1}{4C176B}{white}\blc{18}{2}{5A2D92}{white}\blc{18}{3}{5E4CAA}{white}\blc{18}{4}{6069B7}{white}\blc{18}{5}{6785BE}{white}\blc{18}{6}{789EC2}{blrdark}\blc{18}{7}{96B6C7}{blrdark}\blc{18}{8}{BBCAD1}{blrdark}\blc{18}{9}{DAD8DF}{blrdark}\blc{18}{10}{E0D6D9}{blrdark}\blc{18}{11}{D6C1B4}{blrdark}\blc{18}{12}{CCA389}{blrdark}\blc{18}{13}{C58569}{blrdark}\blc{18}{14}{BA6657}{white}\blc{18}{15}{AA4A50}{white}\blc{18}{16}{943150}{white}\blc{18}{17}{781F4F}{white}\blc{18}{18}{561546}{white}
\blc{19}{1}{4C176B}{white}\blc{19}{2}{5A2B90}{white}\blc{19}{3}{5E48A8}{white}\blc{19}{4}{5F65B5}{white}\blc{19}{5}{657FBD}{white}\blc{19}{6}{7297C1}{blrdark}\blc{19}{7}{8AAEC5}{blrdark}\blc{19}{8}{ACC2CC}{blrdark}\blc{19}{9}{CED3D9}{blrdark}\blc{19}{10}{E2D9E2}{blrdark}\blc{19}{11}{DCCECB}{blrdark}\blc{19}{12}{D1B6A3}{blrdark}\blc{19}{13}{CA9A7D}{blrdark}\blc{19}{14}{C27C63}{blrdark}\blc{19}{15}{B75F55}{white}\blc{19}{16}{A74650}{white}\blc{19}{17}{912F50}{white}\blc{19}{18}{751E4F}{white}\blc{19}{19}{561546}{white}
\blc{20}{1}{4C176B}{white}\blc{20}{2}{592A8F}{white}\blc{20}{3}{5E45A6}{white}\blc{20}{4}{5F60B4}{white}\blc{20}{5}{6379BB}{white}\blc{20}{6}{6E91C0}{white}\blc{20}{7}{82A7C3}{blrdark}\blc{20}{8}{9EBBC9}{blrdark}\blc{20}{9}{C1CDD3}{blrdark}\blc{20}{10}{DBD8DF}{blrdark}\blc{20}{11}{E0D6DA}{blrdark}\blc{20}{12}{D8C5BB}{blrdark}\blc{20}{13}{CEAC94}{blrdark}\blc{20}{14}{C89174}{blrdark}\blc{20}{15}{C0755E}{white}\blc{20}{16}{B45A53}{white}\blc{20}{17}{A44250}{white}\blc{20}{18}{8E2C50}{white}\blc{20}{19}{741E4F}{white}\blc{20}{20}{561546}{white}
\subsection*{$n=9$}
\setlength{\blrslot}{0.245\linewidth}
\noindent%
\blrcard{$4\times9$: \textbf{53.3} \textcolor{gray}{(\sout{56.0}, $-4.8\%$)}}{9}
  {{1,2,3,4,5,6,7,8,9},{5,1,7,3,9,2,6,4,8},{2,7,8,5,4,3,9,1,6},{3,8,5,6,2,9,1,7,4}}%
\hfill%
\blrcard{$5\times9$: \textbf{48.6} \textcolor{gray}{(\sout{52.6}, $-7.6\%$)}}{9}
  {{1,2,3,4,5,6,7,8,9},{9,4,5,2,7,1,8,3,6},{2,6,9,1,8,4,5,7,3},{7,1,4,6,9,3,2,5,8},{4,8,2,7,3,9,6,1,5}}%
\hfill%
\blrcard{$6\times9$: \textbf{40.0} \textcolor{gray}{(\sout{48.0}, $-16.7\%$)}}{9}
  {{1,2,3,4,5,6,7,8,9},{3,8,4,1,7,2,9,5,6},{7,4,5,9,3,1,2,6,8},{4,9,1,8,6,5,3,7,2},{9,3,6,2,4,7,8,1,5},{6,1,7,3,9,8,5,2,4}}%
\hfill%
\blrcard{$7\times9$: \textbf{43.3} \textcolor{gray}{(\sout{51.3}, $-15.6\%$)}}{9}
  {{1,2,3,4,5,6,7,8,9},{6,1,9,3,4,8,5,2,7},{2,9,4,8,6,5,3,7,1},{9,3,5,7,1,2,4,6,8},{8,5,1,9,2,7,6,3,4},{3,8,2,6,7,9,1,4,5},{7,4,8,1,9,3,2,5,6}}%
\par\medskip\noindent%
\blrcard{$8\times9$: \textbf{42.6} \textcolor{gray}{(\sout{53.3}, $-20.0\%$)}}{9}
  {{1,2,3,4,5,6,7,8,9},{8,4,2,9,7,3,5,1,6},{4,5,6,2,9,7,1,3,8},{9,3,7,5,1,4,8,6,2},{7,6,4,3,8,1,2,9,5},{6,9,1,8,4,5,3,2,7},{5,7,8,1,6,2,9,4,3},{3,8,5,6,2,9,4,7,1}}%
\hfill%
\begin{minipage}[t]{\blrslot}\end{minipage}%
\hfill%
\begin{minipage}[t]{\blrslot}\end{minipage}%
\hfill%
\begin{minipage}[t]{\blrslot}\end{minipage}%
\par\medskip
\subsection*{$n=10$}
\setlength{\blrslot}{0.327\linewidth}
\noindent%
\blrcard{$6\times10$: \textbf{60.0} \textcolor{gray}{(\sout{70.0}, $-14.3\%$)}}{10}
  {{1,2,3,4,5,6,7,8,9,10},{8,3,6,9,2,7,5,4,10,1},{3,7,10,5,1,8,9,2,6,4},{5,8,4,3,9,10,2,1,7,6},{9,5,1,6,3,4,10,7,8,2},{7,4,9,1,8,3,6,10,2,5}}%
\hfill%
\blrcard{$7\times10$: \textbf{58.0} \textcolor{gray}{(\sout{72.0}, $-19.4\%$)}}{10}
  {{1,2,3,4,5,6,7,8,9,10},{6,4,7,9,1,2,10,3,8,5},{3,10,4,1,7,8,9,5,6,2},{8,1,9,6,3,5,4,10,2,7},{5,7,1,3,9,10,6,2,4,8},{9,3,2,7,8,4,5,6,10,1},{4,9,5,10,2,1,8,7,3,6}}%
\hfill%
\blrcard{$8\times10$: \textbf{58.6} \textcolor{gray}{(\sout{82.6}, $-29.0\%$)}}{10}
  {{1,2,3,4,5,6,7,8,9,10},{8,6,4,2,9,1,10,3,5,7},{7,1,10,6,8,3,4,9,2,5},{2,7,6,9,4,10,3,5,8,1},{10,8,2,5,3,7,6,4,1,9},{3,9,8,7,6,5,1,2,10,4},{6,5,9,1,10,8,2,7,4,3},{5,4,7,8,1,2,9,10,3,6}}%
\par\medskip\noindent%
\blrcard{$9\times10$: \textbf{64.0} \textcolor{gray}{(\sout{70.0}, $-8.6\%$)}}{10}
  {{1,2,3,4,5,6,7,8,9,10},{7,1,10,9,3,4,8,5,6,2},{9,6,2,7,1,5,3,10,4,8},{6,8,1,5,10,3,9,7,2,4},{10,3,6,2,7,8,4,1,5,9},{4,5,7,10,6,9,2,3,8,1},{3,9,4,6,8,7,1,2,10,5},{2,10,9,8,4,1,5,6,7,3},{8,7,5,3,9,2,10,4,1,6}}%
\hfill%
\begin{minipage}[t]{\blrslot}\end{minipage}%
\hfill%
\begin{minipage}[t]{\blrslot}\end{minipage}%
\par\medskip
\subsection*{$n=11$}
\setlength{\blrslot}{0.327\linewidth}
\noindent%
\blrcard{$4\times11$: \textbf{104.0} \textcolor{gray}{(\sout{112.0}, $-7.1\%$)}}{11}
  {{1,2,3,4,5,6,7,8,9,10,11},{4,3,10,11,6,8,5,1,2,9,7},{9,6,4,2,11,1,10,7,8,3,5},{8,4,5,9,1,10,2,11,3,7,6}}%
\hfill%
\blrcard{$5\times11$: \textbf{94.0} \textcolor{gray}{(\sout{116.0}, $-19.0\%$)}}{11}
  {{1,2,3,4,5,6,7,8,9,10,11},{8,3,2,11,7,4,6,10,1,5,9},{3,10,7,5,1,11,8,6,2,9,4},{7,1,8,9,2,10,3,5,4,11,6},{10,8,4,1,6,3,9,11,7,2,5}}%
\hfill%
\blrcard{$6\times11$: \textbf{88.0} \textcolor{gray}{(\sout{108.0}, $-18.5\%$)}}{11}
  {{1,2,3,4,5,6,7,8,9,10,11},{11,4,6,1,7,8,10,3,2,5,9},{8,3,1,11,9,5,6,4,10,7,2},{3,6,7,9,1,11,2,10,4,8,5},{4,7,9,8,2,3,11,5,6,1,10},{7,5,11,3,10,4,1,9,8,2,6}}%
\par\medskip\noindent%
\blrcard{$7\times11$: \textbf{84.0} \textcolor{gray}{(\sout{118.0}, $-28.8\%$)}}{11}
  {{1,2,3,4,5,6,7,8,9,10,11},{3,5,10,11,1,4,6,7,2,9,8},{6,11,4,3,9,8,1,5,10,2,7},{9,1,7,6,10,3,5,4,8,11,2},{5,6,9,2,11,7,3,10,1,8,4},{8,3,6,10,2,1,11,9,7,4,5},{10,4,2,9,6,5,8,1,11,7,3}}%
\hfill%
\blrcard{$8\times11$: \textbf{78.0} \textcolor{gray}{(\sout{114.0}, $-31.6\%$)}}{11}
  {{1,2,3,4,5,6,7,8,9,10,11},{2,8,11,5,6,7,4,10,1,9,3},{8,4,7,6,3,9,5,2,11,1,10},{6,9,2,10,7,1,8,3,5,11,4},{10,7,5,3,2,8,9,11,4,6,1},{3,11,6,7,1,2,10,4,8,5,9},{7,5,1,9,11,4,2,6,10,3,8},{5,6,10,1,8,3,11,9,7,4,2}}%
\hfill%
\blrcard{$9\times11$: \textbf{84.0} \textcolor{gray}{(\sout{120.0}, $-30.0\%$)}}{11}
  {{1,2,3,4,5,6,7,8,9,10,11},{7,3,6,10,8,1,9,5,4,11,2},{6,8,2,5,7,9,11,10,3,4,1},{3,11,7,2,9,8,4,1,10,6,5},{8,4,5,11,3,10,6,2,1,9,7},{5,7,10,1,11,2,8,4,6,3,9},{11,6,1,9,4,7,5,3,8,2,10},{9,5,8,3,1,11,10,6,2,7,4},{4,10,9,6,2,5,3,11,7,1,8}}%
\par\medskip\noindent%
\blrcard{$10\times11$: \textbf{100.0} \textcolor{gray}{(\sout{110.0}, $-9.1\%$)}}{11}
  {{1,2,3,4,5,6,7,8,9,10,11},{5,7,1,2,10,11,9,4,3,6,8},{3,8,5,10,11,2,4,6,7,9,1},{8,1,7,11,9,3,5,2,6,4,10},{7,6,11,3,2,10,1,9,5,8,4},{6,10,8,1,4,5,11,7,2,3,9},{11,5,4,9,6,1,3,10,8,2,7},{10,4,9,7,3,8,2,5,1,11,6},{2,9,6,5,8,7,10,11,4,1,3},{4,11,2,8,1,9,6,3,10,7,5}}%
\hfill%
\begin{minipage}[t]{\blrslot}\end{minipage}%
\hfill%
\begin{minipage}[t]{\blrslot}\end{minipage}%
\par\medskip
\subsection*{$n=12$}
\setlength{\blrslot}{0.327\linewidth}
\noindent%
\blrcard{$3\times12$: \textbf{152.0} \textcolor{gray}{(\sout{154.0}, $-1.3\%$)}}{12}
  {{1,2,3,4,5,6,7,8,9,10,11,12},{5,10,9,8,2,1,11,12,6,3,4,7},{6,9,10,7,1,2,12,11,5,4,3,8}}%
\hfill%
\blrcard{$4\times12$: \textbf{149.3} \textcolor{gray}{(\sout{154.6}, $-3.4\%$)}}{12}
  {{1,2,3,4,5,6,7,8,9,10,11,12},{4,9,11,1,6,5,8,7,2,12,3,10},{6,3,5,9,12,10,4,1,8,11,2,7},{5,11,6,2,10,12,1,4,7,3,9,8}}%
\hfill%
\blrcard{$5\times12$: \textbf{140.0} \textcolor{gray}{(\sout{145.3}, $-3.7\%$)}}{12}
  {{1,2,3,4,5,6,7,8,9,10,11,12},{11,8,6,3,1,9,4,12,10,7,5,2},{9,6,5,12,3,11,2,10,1,8,7,4},{3,7,11,5,4,12,1,9,8,2,6,10},{8,3,12,2,9,7,6,4,11,1,10,5}}%
\par\medskip\noindent%
\blrcard{$7\times12$: \textbf{121.3} \textcolor{gray}{(\sout{164.0}, $-26.0\%$)}}{12}
  {{1,2,3,4,5,6,7,8,9,10,11,12},{3,11,7,8,2,4,9,10,6,12,1,5},{8,1,9,2,7,12,3,5,10,4,6,11},{9,3,10,5,8,11,1,6,2,7,12,4},{2,9,6,11,12,7,5,3,1,8,4,10},{10,7,1,6,9,3,12,4,8,11,5,2},{7,5,4,9,11,1,10,2,12,3,8,6}}%
\hfill%
\blrcard{$8\times12$: \textbf{116.6} \textcolor{gray}{(\sout{148.0}, $-21.2\%$)}}{12}
  {{1,2,3,4,5,6,7,8,9,10,11,12},{7,3,9,11,1,12,6,5,4,2,10,8},{3,5,10,6,12,2,9,7,11,4,8,1},{10,1,6,9,7,5,4,12,2,8,3,11},{8,6,12,3,4,9,1,2,10,11,5,7},{9,4,5,8,10,3,11,1,6,12,7,2},{6,11,4,7,2,10,3,9,8,1,12,5},{4,12,7,10,3,1,8,11,5,6,2,9}}%
\hfill%
\blrcard{$9\times12$: \textbf{106.0} \textcolor{gray}{(\sout{154.0}, $-31.2\%$)}}{12}
  {{1,2,3,4,5,6,7,8,9,10,11,12},{9,4,12,5,1,7,8,10,3,2,6,11},{2,12,10,1,7,5,6,11,8,9,4,3},{7,9,2,8,3,12,5,6,11,4,10,1},{11,1,7,3,8,4,10,12,5,6,2,9},{8,10,5,2,4,11,9,1,7,12,3,6},{10,7,4,6,9,2,11,3,12,1,5,8},{6,8,1,9,10,3,12,4,2,11,7,5},{5,3,9,10,11,1,2,7,6,8,12,4}}%
\par\medskip\noindent%
\blrcard{$10\times12$: \textbf{124.0} \textcolor{gray}{(\sout{174.6}, $-29.0\%$)}}{12}
  {{1,2,3,4,5,6,7,8,9,10,11,12},{12,1,6,8,2,10,9,11,5,3,7,4},{4,8,12,3,9,1,5,10,11,2,6,7},{8,6,7,9,3,5,10,2,12,11,4,1},{5,11,9,1,6,7,12,3,4,8,10,2},{9,7,4,2,10,11,8,6,1,12,3,5},{6,4,5,10,12,9,2,7,3,1,8,11},{10,5,8,7,1,4,11,12,2,6,9,3},{7,12,2,5,11,8,3,4,6,9,1,10},{3,10,11,6,7,12,4,1,8,5,2,9}}%
\hfill%
\blrcard{$11\times12$: \textbf{136.6} \textcolor{gray}{(\sout{147.3}, $-7.2\%$)}}{12}
  {{1,2,3,4,5,6,7,8,9,10,11,12},{3,7,4,10,9,12,1,6,5,11,2,8},{2,4,12,11,6,9,5,10,3,8,1,7},{11,10,5,1,4,8,6,7,12,3,9,2},{8,12,10,3,1,5,4,11,2,7,6,9},{9,3,11,5,7,4,8,12,6,2,10,1},{7,11,1,2,12,10,9,3,8,4,5,6},{5,9,2,8,10,7,12,4,1,6,3,11},{6,1,9,12,8,11,3,2,7,5,4,10},{12,5,7,6,3,2,10,1,11,9,8,4},{10,6,8,7,11,3,2,9,4,1,12,5}}%
\hfill%
\begin{minipage}[t]{\blrslot}\end{minipage}%
\par\medskip
\subsection*{$n=13$}
\setlength{\blrslot}{0.490\linewidth}
\noindent%
\blrcard{$4\times13$: \textbf{201.3} \textcolor{gray}{(\sout{211.3}, $-4.7\%$)}}{13}
  {{1,2,3,4,5,6,7,8,9,10,11,12,13},{7,4,11,5,9,1,12,13,3,2,8,10,6},{3,11,10,1,7,8,5,6,13,12,4,2,9},{5,10,2,7,13,12,3,9,1,11,6,4,8}}%
\hfill%
\blrcard{$5\times13$: \textbf{191.3} \textcolor{gray}{(\sout{210.6}, $-9.2\%$)}}{13}
  {{1,2,3,4,5,6,7,8,9,10,11,12,13},{2,9,11,1,8,12,3,7,6,5,13,4,10},{6,11,7,5,12,2,1,13,10,3,4,9,8},{7,3,6,9,13,11,10,1,2,4,12,8,5},{12,4,9,6,7,1,2,10,13,8,5,11,3}}%
\par\medskip\noindent%
\blrcard{$6\times13$: \textbf{166.0} \textcolor{gray}{(\sout{204.0}, $-18.6\%$)}}{13}
  {{1,2,3,4,5,6,7,8,9,10,11,12,13},{11,1,7,6,10,12,2,4,13,8,5,3,9},{8,10,1,9,2,11,5,12,3,7,4,13,6},{10,3,6,5,12,1,9,13,7,2,8,4,11},{2,6,9,11,13,10,4,5,1,3,7,8,12},{7,5,10,2,3,13,8,11,6,1,12,9,4}}%
\hfill%
\blrcard{$7\times13$: \textbf{164.0} \textcolor{gray}{(\sout{203.3}, $-19.3\%$)}}{13}
  {{1,2,3,4,5,6,7,8,9,10,11,12,13},{3,8,6,10,1,12,11,9,2,13,4,5,7},{7,1,11,6,13,9,3,4,10,5,8,2,12},{13,6,8,2,11,4,10,1,7,12,5,9,3},{5,10,1,13,8,2,9,11,3,7,12,4,6},{11,4,5,8,3,13,1,7,12,2,10,6,9},{8,9,4,1,12,5,6,13,11,3,2,7,10}}%
\par\medskip\noindent%
\blrcard{$8\times13$: \textbf{160.0} \textcolor{gray}{(\sout{220.6}, $-27.5\%$)}}{13}
  {{1,2,3,4,5,6,7,8,9,10,11,12,13},{13,1,5,10,6,4,3,9,11,8,7,2,12},{10,4,11,6,7,2,12,13,1,5,8,9,3},{9,10,1,12,2,5,8,6,7,4,13,3,11},{11,5,8,2,10,3,13,4,12,9,6,7,1},{4,9,7,5,11,13,2,10,3,1,12,8,6},{2,6,13,9,8,11,4,1,5,12,3,10,7},{5,12,6,11,3,9,10,2,13,7,1,4,8}}%
\hfill%
\blrcard{$9\times13$: \textbf{152.0} \textcolor{gray}{(\sout{214.0}, $-29.0\%$)}}{13}
  {{1,2,3,4,5,6,7,8,9,10,11,12,13},{3,6,11,13,8,4,9,7,12,2,1,5,10},{8,13,5,7,12,2,3,11,1,4,6,10,9},{9,1,8,12,3,11,2,5,10,6,13,7,4},{5,11,1,3,7,10,4,9,13,12,8,2,6},{13,3,10,2,9,5,8,4,7,11,12,6,1},{10,8,7,1,6,3,12,2,11,13,4,9,5},{2,7,9,6,11,13,1,10,5,8,3,4,12},{11,4,2,8,10,1,13,12,6,5,9,3,7}}%
\par\medskip\noindent%
\blrcard{$10\times13$: \textbf{154.6} \textcolor{gray}{(\sout{240.6}, $-35.7\%$)}}{13}
  {{1,2,3,4,5,6,7,8,9,10,11,12,13},{7,3,12,9,8,10,11,1,2,5,6,13,4},{8,11,5,13,3,7,1,4,10,2,12,9,6},{4,12,2,11,7,5,9,6,8,13,1,3,10},{12,1,8,5,4,9,6,13,11,3,10,2,7},{6,7,4,8,13,12,3,2,1,11,5,10,9},{2,5,13,12,10,3,8,9,6,7,4,11,1},{10,8,6,2,1,11,12,7,13,4,9,5,3},{9,13,1,7,2,8,10,5,4,12,3,6,11},{5,10,7,6,12,1,13,11,3,9,2,4,8}}%
\hfill%
\blrcard{$11\times13$: \textbf{168.0} \textcolor{gray}{(\sout{224.0}, $-25.0\%$)}}{13}
  {{1,2,3,4,5,6,7,8,9,10,11,12,13},{12,5,10,6,7,2,1,3,13,11,9,4,8},{3,11,12,5,8,10,2,4,1,9,7,13,6},{10,6,4,3,13,11,8,2,12,7,1,5,9},{4,12,9,7,2,3,11,6,10,1,13,8,5},{11,9,6,13,1,5,3,12,4,8,2,10,7},{13,3,7,8,9,1,10,5,11,12,4,6,2},{6,8,2,9,3,12,13,7,5,4,10,11,1},{8,7,11,1,6,4,12,13,2,3,5,9,10},{5,4,13,2,11,7,9,10,8,6,12,1,3},{2,13,1,12,10,8,4,9,6,5,3,7,11}}%
\par\medskip\noindent%
\blrcard{$12\times13$: \textbf{166.0} \textcolor{gray}{(\sout{274.0}, $-39.4\%$)}}{13}
  {{1,2,3,4,5,6,7,8,9,10,11,12,13},{12,8,7,2,10,1,6,13,3,9,4,5,11},{4,12,10,3,13,2,11,5,8,6,1,7,9},{9,11,2,7,12,4,5,3,13,1,6,8,10},{2,13,9,8,6,5,10,12,4,7,3,11,1},{3,9,1,13,8,12,4,7,10,11,5,6,2},{8,7,13,5,11,3,2,10,1,4,12,9,6},{11,1,12,6,2,13,8,4,5,3,9,10,7},{5,10,4,1,9,8,12,11,6,13,7,2,3},{13,4,6,10,7,11,1,9,2,5,8,3,12},{7,6,5,12,3,9,13,1,11,2,10,4,8},{6,3,8,11,4,10,9,2,7,12,13,1,5}}%
\hfill%
\begin{minipage}[t]{\blrslot}\end{minipage}%
\par\medskip
\subsection*{$k=2$, $14\le n\le 20$}
\setlength{\blrslot}{0.490\linewidth}
\noindent%
\blrcard{$2\times14$: \textbf{268.0}}{14}
  {{1,2,3,4,5,6,7,8,9,10,11,12,13,14},{7,8,2,11,1,12,13,4,5,6,14,3,9,10}}%
\hfill%
\blrcard{$2\times15$: \textbf{344.6}}{15}
  {{1,2,3,4,5,6,7,8,9,10,11,12,13,14,15},{2,12,10,8,6,5,15,4,14,3,13,1,11,9,7}}%
\par\medskip\noindent%
\blrcard{$2\times16$: \textbf{422.0}}{16}
  {{1,2,3,4,5,6,7,8,9,10,11,12,13,14,15,16},{9,11,12,2,3,15,5,16,6,7,8,10,14,1,13,4}}%
\hfill%
\blrcard{$2\times17$: \textbf{498.0}}{17}
  {{1,2,3,4,5,6,7,8,9,10,11,12,13,14,15,16,17},{13,15,4,17,2,7,8,9,10,11,12,1,14,3,16,5,6}}%
\par\medskip\noindent%
\blrcard{$2\times18$: \textbf{610.6}}{18}
  {{1,2,3,4,5,6,7,8,9,10,11,12,13,14,15,16,17,18},{6,4,16,2,14,1,12,11,10,9,8,7,18,5,17,3,15,13}}%
\hfill%
\blrcard{$2\times19$: \textbf{722.6}}{19}
  {{1,2,3,4,5,6,7,8,9,10,11,12,13,14,15,16,17,18,19},{6,5,17,16,2,1,13,12,11,9,10,8,7,19,18,4,3,15,14}}%
\par\medskip\noindent%
\blrcard{$2\times20$: \textbf{832.0}}{20}
  {{1,2,3,4,5,6,7,8,9,10,11,12,13,14,15,16,17,18,19,20},{4,3,16,1,14,19,12,11,10,9,8,7,20,5,18,17,2,15,6,13}}%
\hfill%
\begin{minipage}[t]{\blrslot}\end{minipage}%
\par\medskip
\endgroup